\documentclass[fleqn,usenatbib]{mnras}

\usepackage{newtxtext,newtxmath}

\usepackage[T1]{fontenc}

\DeclareRobustCommand{\VAN}[3]{#2}
\let\VANthebibliography\thebibliography
\def\thebibliography{\DeclareRobustCommand{\VAN}[3]{##3}\VANthebibliography}

\usepackage{graphicx}	
\usepackage{amsmath}	
\usepackage{comment}
\usepackage{subfigure}
\usepackage{siunitx} 
\usepackage{xcolor} 
\usepackage[flushleft]{threeparttable} 
\usepackage{longtable,booktabs}
\usepackage[online]{threeparttablex}
\usepackage{comment}
\usepackage{pdflscape} 
\usepackage{afterpage} 
\usepackage[labelfont=bf, font=small]{caption} 
\usepackage{adjustbox}
\usepackage{soul} 

\newcommand\hi{$\textrm{H}\,\scriptstyle\mathrm{I}$} 
\newcommand\HI{$\textbf{H}\,\scriptstyle\mathbf{I}$}
\newcommand\kms{$\rm km\,s^{-1}$}
\newcommand\mJy{$\rm mJy\,beam^{-1}$}
\newcommand{\msun}{M$_\odot$}
\defcitealias{Kraan-Korteweg2017}{KK2017}

\title[A blind MeerKAT \hi{} Survey of the Vela Supercluster]{Revealing hidden structures in the Zone of Avoidance -- a blind MeerKAT \hi{}~~Survey of the Vela Supercluster}

\author[S. H. A. Rajohnson et al.]{
Sambatriniaina H. A. Rajohnson$^{1}$,\thanks{E-mail: aychasam@gmail.com}
Renée C. Kraan-Korteweg$^{1}$,
Bradley S. Frank$^{2,3,4,1}$,
Hao Chen$^{1,5}$,
\newauthor{Lister Staveley-Smith$^{6}$,
Paolo Serra$^{7}$,
Nadia Steyn$^{1,6}$,
Sushma Kurapati$^{1}$,
D. J. Pisano$^{1}$,
Sharmila Goedhart$^{4,8}$}
\\
$^{1}$Department of Astronomy, University of Cape Town, Private Bag X3, Rondebosch 7701, South Africa\\
$^{2}$UK Astronomy Technology Centre, Royal Observatory Edinburgh, Blackford Hill, Edinburgh EH9 3HJ, UK\\
$^{3}$South African Radio Astronomy Observatory (SARAO), 2 Fir Street, Observatory, 7925, South Africa\\
$^{4}$The Inter-University Institute for Data Intensive Astronomy (IDIA), and University of Cape Town, Private Bag X3, Rondebosch, 7701, South Africa\\
$^{5}$Research Center for Intelligent Computing Platforms, Zhejiang Laboratory, Hangzhou 311100, China\\
$^{6}$International Centre for Radio Astronomy Research (ICRAR), The University of Western Australia, 35 Stirling Highway, Australia\\
$^{7}$INAF – Osservatorio Astronomico di Cagliari, Via della Scienza 5, 09047, Selargius, CA, Italy\\
$^{8}$SKAO, 2 Fir Street, Black River Park, Second Floor, Block A, Cape Town, 7925\\
}

\date{Accepted 2024 November 09. Received 2024 November 07; in original form 2024 September 07}

\pubyear{2024}

\begin{document}
\label{firstpage}
\pagerange{\pageref{firstpage}--\pageref{lastpage}}
\maketitle

\begin{abstract}

We conducted the MeerKAT Vela Supercluster survey, named Vela$-$\hi{}, to bridge the gap between the Vela SARAO MeerKAT Galactic Plane Survey (Vela$-$SMGPS, $-2^{\circ} \leq b \leq 1^{\circ}$), and optical and near-infrared spectroscopic observations of the Vela Supercluster (hereafter Vela$-$OPT/NIR) at $|b| \gtrsim 7^{\circ}$. Covering coordinates from $263^{\circ} \leq \ell \leq 284^{\circ}$ and $1^{\circ} \leq b \leq 6.2^{\circ}$ above, and $-6.7^{\circ} \leq b \leq -2^{\circ}$ below the Galactic Plane (GP), we sampled 667 fields spread across an area of ${\sim} \rm 242 ~deg^2$. With a beam size of ${\sim} \ang{;;38} \times \ang{;;31}$, Vela$-$\hi{} achieved a sensitivity of $\langle \rm rms \rangle = 0.74$ \mJy{} at 44.3 \kms{} velocity resolution over ${\sim}$67 hours of observations. We cataloged 719 galaxies, with only 211 (29\%)  previously documented in the literature, primarily through the HIZOA, 2MASX, and WISE databases. Among these known galaxies, only 66 had optical spectroscopic redshift information. We found marginal differences of less than one channel resolution for all galaxies in common between HIZOA and Vela$-$SMGPS, and a mean difference of $70 \pm 15$ \kms{} between optical and \hi{} velocities. 
Combining data from Vela$-$SMGPS, Vela$-$\hi{}, and Vela$-$OPT/NIR confirmed the connection of the Hydra/Antlia filament across the GP and revealed a previously unknown diagonal wall at a heliocentric velocity range of $6500-8000$ \kms{}.  
Vela$-$\hi{} reinforces the connection between the first wall at $18500-20000$ \kms{} and the inner ZOA. The second wall seems to traverse the GP at 
$270^{\circ} \leq \ell \leq 279^{\circ}$, where it appears that both walls intersect, jointly covering the velocity range $18500-21500$ \kms{}. 
\end{abstract}




\section{Introduction}

The recently discovered Vela Supercluster (hereafter VSCL) is of particular interest among the large-scale structures (LSS) bisected by the ZOA because of its close alignment with the apex of the residual bulk flow, 
which is expected to originate from distances beyond $cz \geq 16000$ \kms{} \citep{Springob2016,Scrimgeour2016}, and even outside 20000 \kms{} ($200\,h^{-1}$\,Mpc, \citealt{Carrick2015,Boubel2024}). As part of a wide-area campaign aiming to explore LSS along the entire 360$^{\circ}$ circle of the ZOA and learn more about the overall bulk flow, \citealt{Kraan-Korteweg2017} (\citetalias{Kraan-Korteweg2017}) pursued spectroscopic observations encompassing the region that contains the VSCL. The ${\sim}4200$ new redshifts revealed it as an extended massive overdensity with multibranching filaments \citep{Einasto2011}. Its mass excess was estimated to contribute to ${\sim}50$ \kms{} to the motion of the Local Group (LG). However, optical (OPT) and near-infrared (NIR) data provide hardly any data that allow the detailed mapping of its overdensity and surmised walls because of dust obscuration and high star density at latitudes below $|b| < 5^{\circ}$. However, independent analyses using kinematic models have hinted at the presence of the dense core of VSCL deep within the ZOA \citep{Sorce2017,Courtois2019}.\\

The results of a first extensive exploration of the VSCL using \hi{} data from the Vela-SARAO MeerKAT Galactic Plane survey  (Vela$-$SMGPS) has proven to be highly successful \citep{Rajohnson2024}. The survey spans a region of 90 deg$^2$ along a 3$^{\circ}$ latitude strip centered on $\ell \sim 275^{\circ}$. In total, 843 heavily obscured galaxies were identified. Following that project, we conducted a blind systematic interferometric survey referred to as the MeerKAT Vela Supercluster survey or Vela$-$\hi{} that will close the gap between Vela$-$SMGPS and OPT and NIR spectroscopic surveys. It covers an area of $263^{\circ}\leq \ell \leq 284^{\circ}, |b| \lesssim 6.7^{\circ}$, with its 667 pointings distributed both above and below the GP. The large survey area of ${\sim}242 \rm ~ deg^2$ aims to achieve the following three main objectives:

\emph{(i)} To cover the area that bridges the gap between Vela$-$SMGPS \citep{Rajohnson2024}, and ancillary data in the more transparent surroundings at latitudes $|b| \gtrsim 7^{\circ}$, where data availability is less scarce. These OPT and NIR data include published data from \citetalias{Kraan-Korteweg2017} (AAOmega+2dF+SALT), along with additional redshift data provided by OPTOPUS and 6dF in the Hydra/Antlia region that remain unpublished (Kraan-Korteweg, priv. comm). These data will henceforth be referred to as the Vela$-$OPT/NIR dataset. 

\emph{(ii)} To combine the results obtained from Vela$-$\hi{} with those from Vela$-$SMGPS, encompassing various structures out to 25000 \kms{}, to get a better census of the missing part of the VSCL walls at higher latitudes of the ZOA and to obtain an improved view of the extent of the VSCL and its morphology.

\emph{(iii)} In addition, the Vela$-$\hi{} will provide insight into nearby structures that were previously incompletely mapped or unknown. \\

The paper is organized as follows. We present the observing strategy of Vela$-$\hi{} in Section \ref{sec5:survey_design}. Section \ref{sec5:data_processing} delves into the techniques and steps employed for \hi{} data processing, such as data reduction, mosaic generation, and source identification. The results are presented in Section \ref{sec5:results}, accompanied by an assessment of the reliability of the extracted \hi{} galaxies and a comparison of properties with cross-matched data from other surveys, both \hi{} or multi-wavelength data. In Section \ref{sec5:LSS}, we explore the alignment of the newly discovered galaxies with known LSS and analyze the identified structures in combination with the Vela$-$SMGPS detections and galaxies adjacent to the Vela$-$\hi{} survey region. A comprehensive summary of our findings is then provided in Section \ref{sec5:conclusion}.

Throughout this work, all velocities are in the heliocentric optical convention ($V_{\rm hel} = cz$). We adopt a standard cosmological model with H$_0 = 70$ \kms{} Mpc$^{-1}$, $\Omega{\rm m} = 0.3$, and $\Omega_{\Lambda} = 0.7$.

\section{Survey design and Observation Strategy}\label{sec5:survey_design}
Vela$-$\hi{} was observed using the full MeerKAT array \citep{Jonas2016} in eight sessions between August 15 and October 30, 2021 (MeerKAT Open-Time Proposal ID: SCI-20210212-SR-01). The observations were carried out with an array of 60 to 63 antennas using the 4k spectral line correlator mode in L-band ($856 - 1712$ MHz bandwidth), with visibilities recorded every 8 seconds. We used the coarse channel spacing of 209 kHz, which corresponds to a rest-frame velocity resolution of ${\sim} 44$ \kms{} at $z = 0$. 
This setup has been found to be sufficient for detecting and extracting \hi{}-signatures of normal spiral galaxies with typical linewidths of $w_{50} > 150$ \kms{} (see our results from the full SMGPS, \citealt{Kurapati2024,Rajohnson2024,Steyn2024}).

\begin{figure*}
    \centering    
    \includegraphics[width=0.8\linewidth]
    {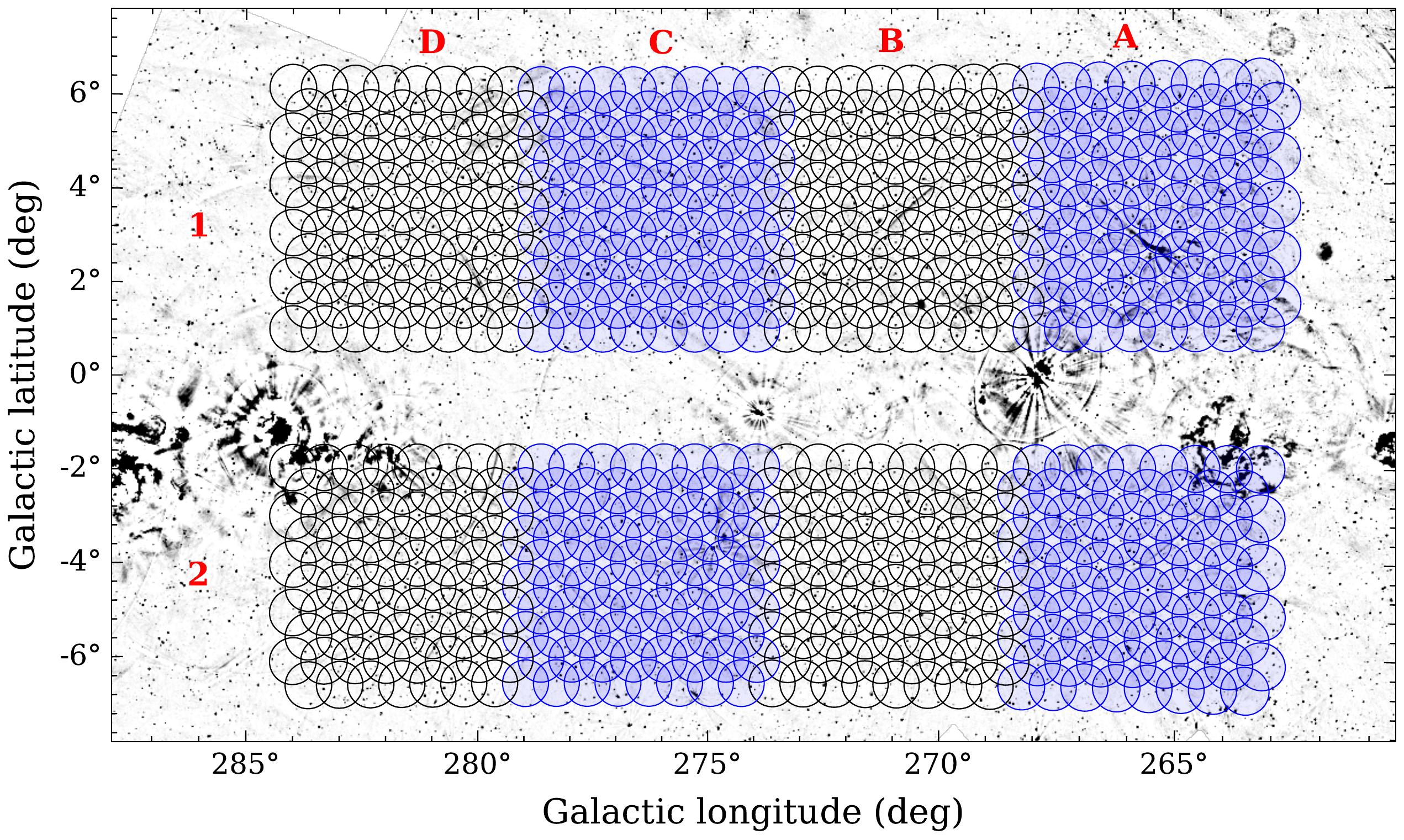}
    \caption{The 242 deg${^2}$ area covered by the MeerKAT Vela Supercluster Survey, overlayed on a grayscale continuum map at 843 MHz from SUMSS (the Sydney University Molonglo Sky Survey, \citealt{Bock1999}). The 667 individual pointings are distributed into 8 observing blocks (A1, A2, B1, B2, C1, C2, D1, and D2) above and below the GP. The gap between the two sets of pointings has been covered by the Vela$-$SMGPS survey \citep{Rajohnson2024}.}
    \label{fig:blocks}
\end{figure*}

Vela$-$\hi{} covers two strips spanning approximately $\Delta l \times \Delta b ~ {\sim} ~ 21^{\circ} \times 6^{\circ}$ on either side of the GP. With the MeerKAT Primary Beam Full Width Half Maximum (FWHM) of ${\sim}1.02^{\circ}$ at 1.4 GHz,
the survey consists of 352 and 315 fields for the strips above and below the GP, respectively, as illustrated in Fig. \ref{fig:blocks}. The area between $-2^{\circ}\leq b \leq 1^{\circ}$ was already surveyed by Vela$-$SMGPS. To ensure uniform sensitivity across the survey area, we used a Nyquist sampling configuration with a spacing of $r = 0.671^{\circ}$.\\

The observations were divided into 8 sessions, referred to as `blocks', each having been observed between $7.6$ to $9$ hours. While the majority of the data collection occurred at night, some data in each block were observed during the day. The data captured during daylight hours were susceptible to solar radio-frequency interference (RFI) and required special flagging strategies (see Appendix \ref{sec5:RFI_sol}).

The strip above the GP was observed in four blocks, denoted by suffix 1, with field prefixes ranging from A to D in increasing longitudes. Each block consists of 88 fields. The strip below the GP was observed in another set of four blocks, labeled with the suffix 2. These blocks contained 80 fields each, except for block D2 (lower left), which consisted of 75 fields. Within a block, each target field was visited twice during a cycle of observation, with each scan lasting 150 seconds. Bandpass and flux calibrators were observed for 3 minutes at the beginning, after the first cycle, and at the end of each run. Gain calibrators were observed for 2 minutes every 25 minutes, repeated in general after 10 fields had been observed. The complete survey was performed in approximately 56 hours of on-source time, with a total of 67.44 hours including overhead. Details of the observations for each block, listed in ascending order of the observed area, are provided in Table \ref{tab:obs}.

\begin{table*}
    \small
	\centering
        \setlength{\tabcolsep}{4pt}
	\caption{Vela$-$\hi{} observing blocks overview}
	\label{tab:obs}
		\begin{tabular}{@{} c c c c c c c c c c @{}}
		    \hline
			\hline
			Obs Date & Block ID & $\ell$ & $b$ & Fields & Track & N$_{\rm ant}$ & Primary calibrators & Secondary calibrators & Notes \\
                (UT start) &  & (deg) & (deg) & & (h) &  & (gaincal) & (fluxcal, bpcal, delaycal) & \\
			\hline
        2021-08-22 & 1629602495 (A1) & $263^{\circ} \leq \ell \leq 269^{\circ}$ & $1^{\circ} \leq b \leq 6.2^{\circ}$ & 88 & 8.88 & 61 & J0828-3731 & J0408-6545 & a \\
        2021-10-01 & 1633050310 (B1) & $269^{\circ} \leq \ell \leq 274^{\circ}$ & $1^{\circ} \leq b \leq 6.2^{\circ}$ & 88 & 8.84 & 61 & J0825-5010 & J0408-6545 & a\\
        2021-10-30 & 1635552081 (C1) & $274^{\circ} \leq \ell \leq 279^{\circ}$ & $1^{\circ} \leq b \leq 6.2^{\circ}$ & 88 & 8.93 & 61 & J0825-5010 & J0408-6545, J1939-6342 & a \\
        2021-09-04 & 1630723298 (D1) & $279^{\circ} \leq \ell \leq 284^{\circ}$ & $1^{\circ} \leq b \leq 6.2^{\circ}$ & 88 & 8.94 & 61 & J0825-5010, J0906-6829  & J0408-6545, J1939-6342 & a, b\\
        2021-08-15 & 1628997934 (A2) & $263^{\circ} \leq \ell \leq 269^{\circ}$ & $-6.7^{\circ} \leq b \leq -2^{\circ}$ & 80 & 8.11 & 60 & J0828-3731 & J0408-6545 & a \\
        2021-10-24 & 1635037887 (B2) & $269^{\circ} \leq \ell \leq 274^{\circ}$ & $-6.7^{\circ} \leq b \leq -2^{\circ}$ & 80 & 8.06 & 63 & J0825-5010 & J0408-6545, J1939-6342 & a \\
        2021-10-02 & 1633134084 (C2) & $274^{\circ} \leq \ell \leq 279^{\circ}$ & $-6.7^{\circ} \leq b \leq -2^{\circ}$ & 80 & 8.09 & 61 & J0825-5010 & J0408-6545 & a\\
        2021-09-12 & 1631417493 (D2) & $279^{\circ} \leq \ell \leq 284^{\circ}$ & $-6.7^{\circ} \leq b \leq -2^{\circ}$ & 75 & 7.59 & 61 & J0906-6829 & J0408-6545 & a, c\\
            \hline
		\end{tabular}
    \begin{tablenotes}
    \item \footnotesize{$^{\rm a}$ This data dataset is affected by the short track bug (see Appendix \ref{sec5:bug}).}
    \item \footnotesize{$^{\rm b}$ M039 (V-pol) shows dips in its power spectrum and will likely need to be flagged. During the last scan, the M058 data were flagged since it did not lock on target.}
    \item \footnotesize{$^{\rm c}$ There was a wind stow event at the end of the observation.}
    \end{tablenotes}
	\end{table*}

\section{\hi{} data processing}\label{sec5:data_processing}

We closely followed the strategy of Vela$-$SMGPS \citep{Rajohnson2024} for data transfer, reduction procedures, software usage, imaging and \hi{} source finding. Modifications were tailored to suit the specific requirements of the Vela$-$\hi{} dataset, as detailed in the upcoming sections. 

\subsection{Data reduction}

As in Vela$-$SMGPS, we only retrieved the frequency range of $1308 - 1430$ MHz with their parallel-hand polarizations (XX, YY) for each observation from the SARAO archive\footnote{https://archive.sarao.ac.za/} for transfer to the \textsc{ilifu} cloud computing facility\footnote{www.ilifu.ac.za}. This frequency range was selected for its flat, mostly RFI-free band, excluding bandpass roll-offs, and its high-redshift end corresponding to $z \simeq 0.08$ ($cz \sim 25000$ \kms{}) over the full extent of VSCL. The Containerised Automated Radio Astronomical Calibration CARACal 
\citep{Jozsa2020} was used for reducing the Vela$-$\hi{} observations. The pipeline that was executed on each dataset, involved several steps such as automatic flagging, cross-calibration, self-calibration, Doppler-tracking correction to the barycentric frame, continuum subtraction, imaging, and mosaicking. Unlike Vela$-$SMGPS, where a single \textsc{ilifu} node processed a block, handling an entire block of 88 fields in Vela$-$\hi{} is computationally expensive, and will take weeks to complete. To speed up the process, we used ten nodes for parallel processing the data, each managing eight fields, reducing the processing time to around three to four days per block.

Following cross-calibration, the pipeline generated diagnostic plots of the calibrators and target using \textsc{shadeMS} \citep{Smirnov2022}. 
This enabled a swift assessment of the quality of calibrated visibilities and helped identify any remaining RFI for flagging. We encountered two distinct issues during the inspection of cross-calibrated visibilities' diagnostic plots: short-track bugs and solar interference (see Appendices \ref{sec5:bug} and \ref{sec5:RFI_sol}, respectively). Despite setting \textsc{sunblocker}\footnote{https://github.com/gigjozsa/sunblocker} to \textsc{true} during the imaging step, the latter issue persisted. This led us to incorporate an additional manual flagging step (see residual RFI flagging procedures in Appendix \ref{sec5:RFI}) prior to the automatic flagging procedure in the pipeline, which however required a complete re-run of the reduction process.

\subsection{Imaging and mosaicking}\label{sec5:mosaicking}

We employed \textsc{WSClean} for imaging, with a pixel size of $\ang{;;3}$, Briggs $r = 0$, and a UV-taper of $\ang{;;15}$. This produced a total of 667 \hi{} cubes, each composed of 570 channels (ranging from 1309 MHz to 1428.49 MHz), with a width of 210 kHz after Doppler correction. This setup was made to ensure that the cubes have the same zero-point velocity as those for Vela$-$SMGPS, corresponding to a rest-frame velocity resolution of 44.3 \kms{} at $z = 0$ and 46.9 \kms{} at the VSCL distance (i.e., $cz \sim 18000$ \kms{}). After some adjustments, including primary beam (PB) correction and resizing the radii of the \hi{} cubes to a normalized PB sensitivity cut-off of 0.2 (equivalent to $0.8^{\circ}$), the partially overlapping \hi{} cubes were prepared for mosaicking. 

\begin{figure}
    \centering    
    \includegraphics[width=\linewidth]{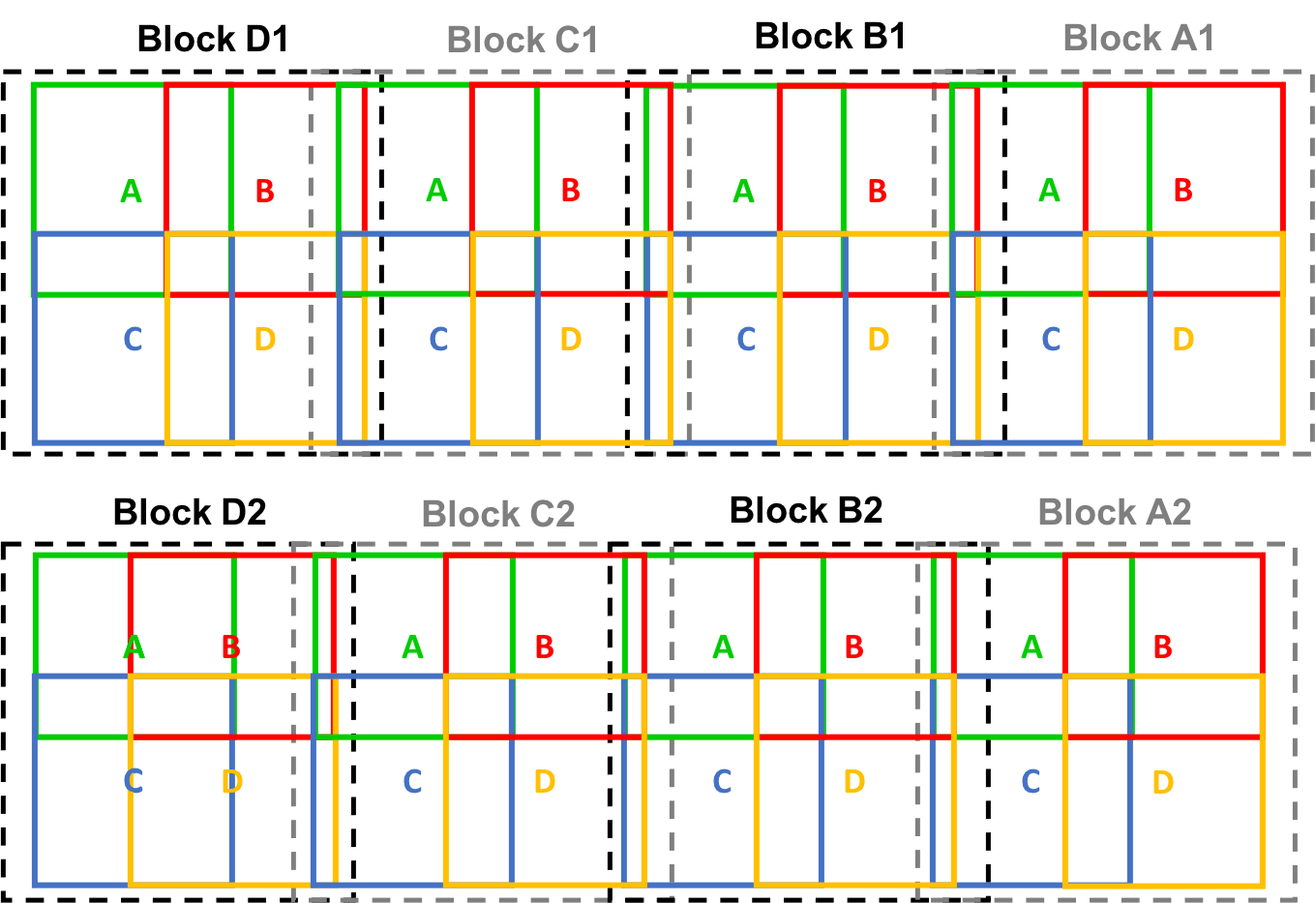}
    \caption{Mosaicking configuration of Vela$-$\hi{}. Each block, indicated by black or grey dashed rectangles, is subdivided into four partially overlapping tiles labeled A to D.}
    \label{fig:blocks_mosaic}
\end{figure}

The \hi{} cubes were extracted per pointing. The eight observing blocks of 80 to 88 cubes, were divided into four mosaics labeled A to D. Above the GP, mosaics were constructed of 27 contiguous \hi{} cubes. For the blocks below the GP, the initial two mosaics (with suffixes A and B) were composed of 22 cubes, while the remaining two (C and D) contained 27 cubes each. 
This arrangement resulted in an on-sky coverage of approximately $3.5^{\circ} \times 4^{\circ}$ or $3.5^{\circ} \times 3.5^{\circ}$ per mosaic, with an overlap of $\Delta \ell = \Delta b \sim 1.5^{\circ}$ (except $\Delta \ell \sim 2^{\circ}$ for D2) for mosaics within the same block, and $\Delta \ell \sim 0.75^{\circ}$ overlap between two mosaics from two adjacent blocks. These overlapping regions were essential for capturing extended sources that might be detected at the edge of the mosaic and were used at the same time for conducting an internal data quality assessment based on the properties of detected galaxies in these regions (see Section \ref{sec5:dupl_analysis}). The overall mosaic layout is illustrated in Fig. \ref{fig:blocks_mosaic}. \\

In total, 32 mosaics were produced, comprising 16 consecutive ones above and below the GP, respectively. These mosaics were generated from individual fields with rms values ranging from 0.88 to 2.57 \mJy{}. The Nyquist sampling setup of the survey has provided us with a nearly uniform central region covering approximately $3^{\circ} \times 3.5^{\circ}$ and $3^{\circ} \times 3^{\circ}$ (utilized for source finding regions). Additionally, it also facilitated a twofold increase in sensitivity after mosaicking compared to an individual field, as illustrated in Fig. \ref{fig:rms_variations} of Appendix \ref{sec5:RFI}. The figure showcases three subfigures presenting the rms variations per field before solar RFI removal, after solar RFI removal, and after mosaicking, respectively.\\

\begin{figure}
    \centering    
    \includegraphics[width=\linewidth]{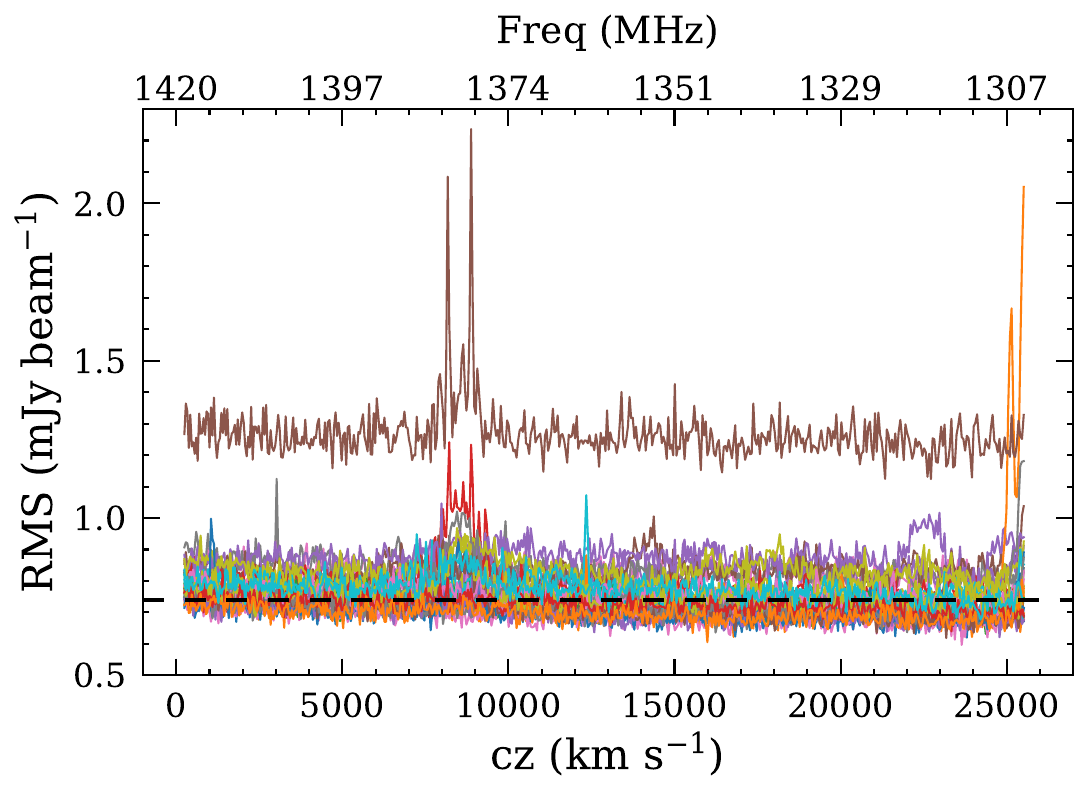}
    \caption{Rms versus velocity in \kms{} as measured through the centers of the 32 mosaics. Interference from Global Positioning
Satellites can be seen in the frequency band 1375 – 1387 MHz (${\sim} 8500$ \kms{}). The average rms noise of 0.74 \mJy\ is shown by the horizontal dashed lines when excluding the high noise from the mosaic A2B (brown line), which is considerably higher than other mosaics and shows stronger fluctuations as well.}
    \label{fig:rms_mosaic}
\end{figure}

The spectral baselines of the central regions of the mosaics generally remain flat with rms noise ranging from 0.68 to 1.24 \mJy{}, as displayed in Fig. \ref{fig:rms_mosaic}. Only for a few cases do we observe increased noise levels towards the edge of the mosaic’s spectral axis ($cz > 24000$ \kms{}). Additionally, residual RFI spikes from Global Positioning Satellites (GPS), i.e., around 1375$-$1387 MHz (${\sim}8500$ \kms{}) appear in only 9 out of 32 mosaics. This leaves a nearly uniform noise variation with no significant noise peaks when considering the remaining 23 mosaics, indicating that RFI has minimal effects on the data, which will not impact the interpretation of LSS in Section \ref{sec5:LSS}. We excluded the high-noise region from mosaic A2B from the subsequent assessment and analysis of the survey. This region shows nearly double the noise compared to other mosaics (see the brown line in Fig. \ref{fig:rms_mosaic}) which may be due to the presence of the Vela Supernova remnant ($\ell \sim 263.93^{\circ}, b \sim -3.36^{\circ}$, \citealt{Green2009}). Exclusion of the A2B mosaic results in an average noise value of 0.74 \mJy{} for the Vela$-$\hi{} survey. For a source detected at 5$\sigma$ and with 200 \kms{} linewidth, this translates to an \hi{} mass limit of 5$\times 10^9$ \msun\ and 7.1$\times 10^9$ \msun\ at $cz \sim 17500$ \kms{} (Wall 1) and $\sim 21000$ \kms{} (Wall 2), respectively.\\

The mosaic beam size is determined as the median value from the individual beams of each input cube. Compared to this median mosaic beam area, we observe a scatter of approximately 11\% in the beam areas per channel and per cube.  
An average beam size of $\ang{;;38} \times \ang{;;31}$
is derived for Vela$-$\hi{}, slightly larger than for SMGPS. This corresponds to a spatial resolution of about $47 \times 39$
kpc at the VSCL distance ($cz \sim 18000$ \kms{}), adequate for our scientific objectives. For comparison, the mosaicking parameters for the two systematic \hi\ surveys, Vela$-$SMGPS and Vela$-$\hi, are provided in Table \ref{tab:survey_comparison}.

\begin{table}
    \small
	\centering
        \setlength{\tabcolsep}{3pt}
	\caption{Survey parameters' comparison of Vela$-$\hi{} and Vela$-$SMGPS}
	\label{tab:survey_comparison}
		\begin{tabular}{l c c}
		    \hline
			\hline
			  Parameter& Vela$-$\hi\ & Vela$-$SMGPS\\
			\hline
          Date of observations & Aug $-$ Oct 2021& Dec 2018 $-$ Nov 2019\\
          Sky coverage& ${\sim}$242 deg$^2$& ${\sim}$90 deg$^2$\\
          Effective area ($\ell$) & $263^{\circ} \leq \ell \leq 284^{\circ}$ & $260^{\circ} \leq \ell \leq 290^{\circ}$ \\
          Effective area ($b$) & $1^{\circ} \leq b \leq 6.2^{\circ}$, & $-2^{\circ} \leq b \leq 1^{\circ}$ \\
           & $-6.7^{\circ} \leq b \leq -2^{\circ}$ & \\
          Number of Mosaics& 32& 10\\
          Integration time$^{\rm a}$& ${\sim}$300 s& ${\sim}$3600 s\\
          Velocity resolution$^{\rm b}$& 44.3 \kms{}& 44.3 \kms{}\\
          Velocity range$^{\rm c}$& $< 25000$ \kms{}& $< 25000$ \kms{}\\
          Beam size& $\ang{;;38} \times \ang{;;31} (\pm \ang{;;3})$& $\ang{;;30} \times \ang{;;26} (\pm \ang{;;1})$\\
          Measured rms & 0.68$-$1.24 \mJy{} & 0.29$-$0.56 \mJy{}\\
          Mean rms & 0.74 \mJy{} & 0.39 \mJy{}\\
            \hline
		\end{tabular}
  \begin{tablenotes}
      \footnotesize
      \item $^{\rm a}$ Integration time per pointing. $^{\rm b}$ at $z = 0$.
      \item $^{\rm c}$ \hi\ source finding velocity range .
    \end{tablenotes}
	\end{table}

\subsection{\hi{} detections}\label{sec5:source_finding}

\begin{figure*}
	\centering
	\subfigure{\includegraphics[width=0.48\linewidth]{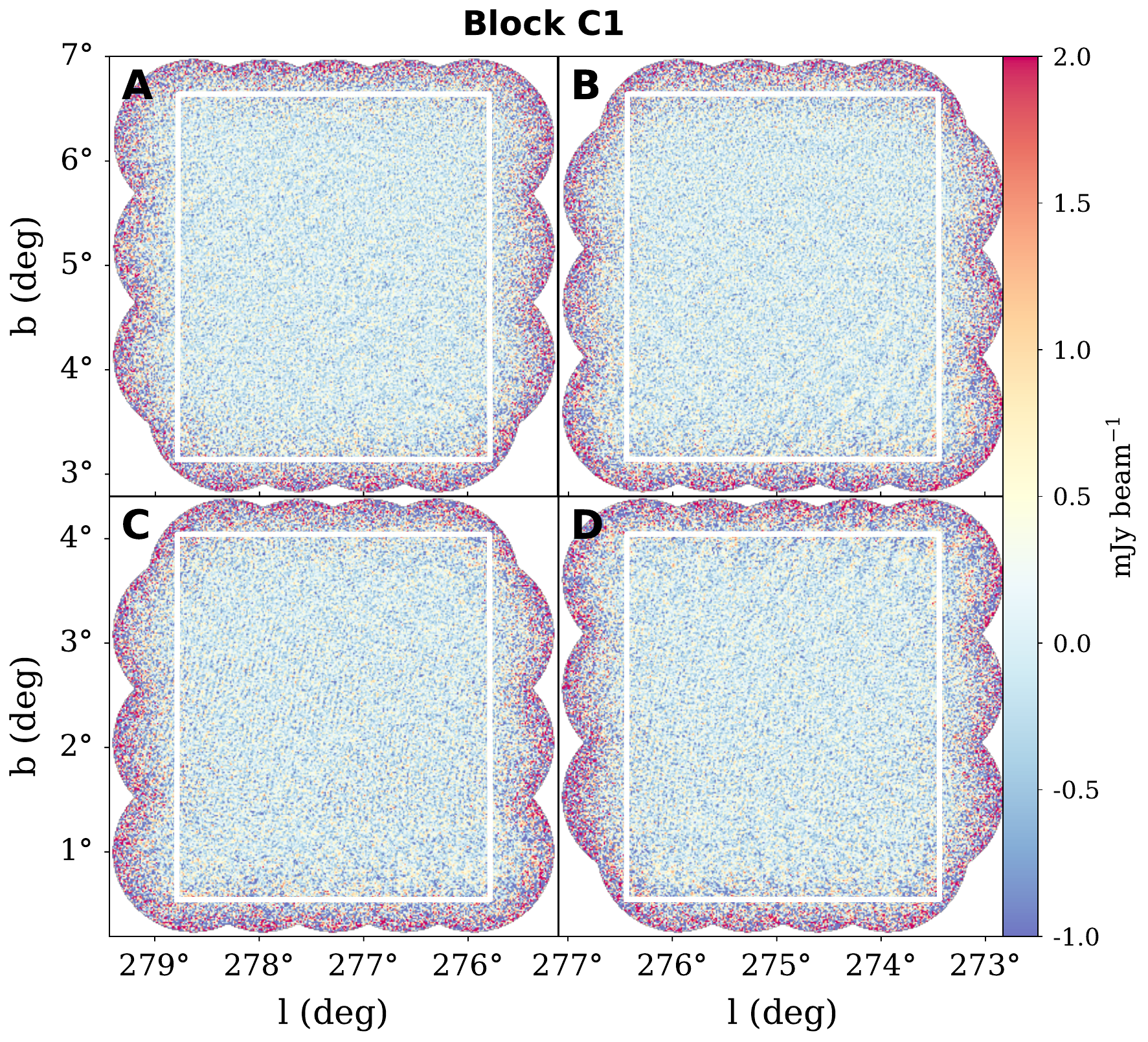}}
	\subfigure{\includegraphics[width=0.49\linewidth]{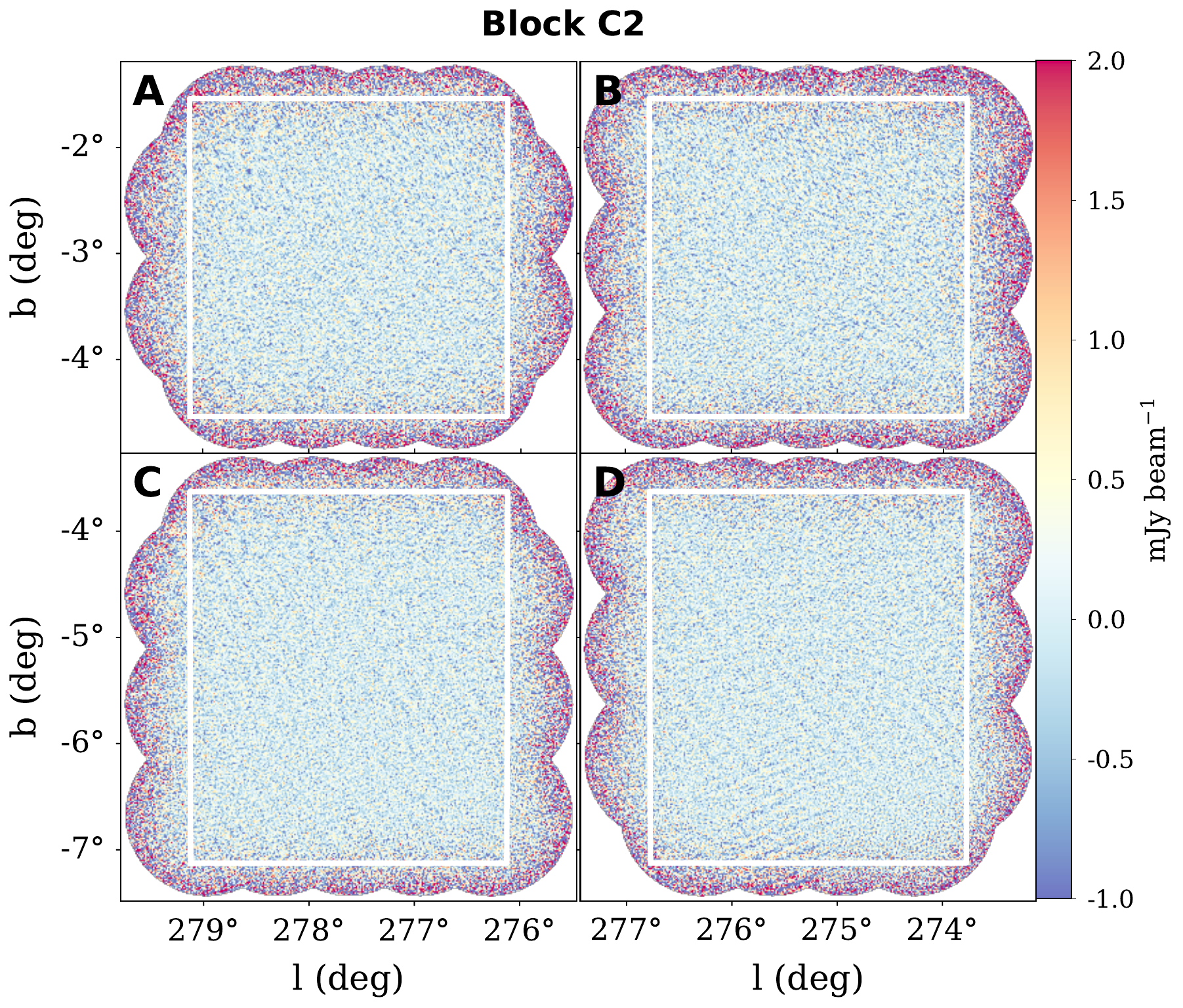}}
	\centering
	\caption{Example of source finding regions for the different mosaic configurations. The top panel illustrates regions of ${\sim} 3.5^{\circ} \times 3^{\circ}$ for a block of four mosaics located above the GP (e.g., block C1). The bottom panel shows two mosaics with ${\sim} 3^{\circ} \times 3^{\circ}$ regions (top), and the other two with ${\sim} 3.5^{\circ} \times 3^{\circ}$ (bottom) for a block located below the GP (e.g., block C2). The white rectangles outline the source finding regions.}
	\label{fig:mosaics_SF}
\end{figure*}

In large surveys like Vela$-$\hi{}, visually identifying sources becomes impractical due to the size of each mosaic, approximately 55 GB with ${\sim} \rm 5150 \rm ~pix \times 5000 \rm ~pix \times 526 \rm ~chans$ voxels. Applying automated source-finding methods is therefore crucial. Following the approach used in Vela$-$SMGPS, we employed the SoFiA-2 algorithm 
\citep{Serra2015,Westmeier2021} to search for \hi{} sources. This process was applied to all mosaics for the velocity range $250 < cz < 25 000$ \kms{}. Two different source-finding regions were defined. The first box encompasses 3600  $\rm pix \times \rm 4200~ pix \times 517~ chans ~({\sim}3.5^{\circ} \times 3^{\circ})$ for the mosaics above the GP and half of the mosaics below the GP. For the remaining half below the GP, we defined mosaics of $\rm 3600~ \rm pix \times 3600~ pix \times 517~ chans ~({\sim}3^{\circ} \times 3^{\circ})$. The source-finding regions are outlined by white rectangles in Fig. \ref{fig:mosaics_SF}, and defined to maximize sensitivity in the central part while avoiding lower sensitivity at the edges of the mosaics. Some overlap between adjacent mosaics is ensured.

Apart from noise normalization through the weighting by noise cubes, we performed an additional continuum subtraction step using a polynomial of order $n=1$ with a threshold of $2\sigma$ to eliminate residual low-level continuum emission. Subsequently, we applied SoFiA's default source-finding algorithm `Smooth and Clip (S+C)'. In comparison to Vela$-$SMGPS, we introduced two additional spatial smoothing kernels of 20 and 30 pixels, equivalent to $\ang{;;60}$ and $\ang{;;90}$ angular resolution smoothing, to account for the larger beam size in our survey. However, the spectral kernels and the two-run thresholds of 3.5 and $4\sigma$ remained consistent with Vela$-$SMGPS. We then merged all detected pixels with a minimum linking size of 14 pixels (slightly larger than one beam) to obtain coherent sources.

For source validation, each SoFiA run produced a catalog of detected sources, including both genuine detections and false positives. To ensure accuracy, three team members independently adjudicated the sources. Each team member carefully reviewed each potential \hi{} source in the catalog by visually inspecting their moment maps and global \hi{} profiles. A `solid' detection shows a clearly defined \hi{} disc in the mom-0 maps, clear rotation across two or more channels in the mom-1 maps, and a distinct peak over a flat baseline in the spectrum. Detections are classified as `possible' if they meet some, but not all, of these criteria; for example, if they exhibit a low-velocity gradient or lack clear emission due to wiggly baselines. Additionally, faint sources smaller than 1.5 times the beam size with a signal-to-noise ratio (SNR) $\leq 6$ were conservatively classified as `possible'. Sources that did not meet these criteria, such as those with rotation over more than 20 channels or high-redshift dwarfs, were removed from the catalog. 
Additional details on this adjudication process can be found in \cite{Rajohnson2024}. The final results of the panels describing the source (i.e., mom-0, mom-1, global \hi{} profile) are presented in a galaxy atlas available in the following Zenodo repository: \url{https://doi.org/10.5281/zenodo.12522807}. 

For source parameterization, SoFiA automatically generates \hi{} parameters in default units of the mosaic, including the central position in Galactic coordinates, central frequencies in Hz, integrated fluxes in Jy Hz, and linewidths in Hz. We applied the rest-frame velocity convention, which was the necessary conversion to obtain the systemic heliocentric velocity in \kms{}, integrated flux in Jy \kms{}, and linewidths in \kms{}, enabling us to derive the \hi{} masses. The uncertainties in the fluxes were calculated by averaging the standard deviations of the integrated fluxes measured from four emission-free regions, each the same size as the source mask and located around the \hi{} source. For full details of the parameterization and error estimation, please refer to Section 3.3 of \cite{Rajohnson2024}.

\section{Results}\label{sec5:results}

We identified a total of 719 \hi\ sources. The complete catalog, along with their derived \hi{} parameters, is available in Table \ref{tab:cat} of Appendix \ref{sec5:full_cat}. In this section, we present the findings and comparative assessment of \hi\ sources in the Vela$-$\hi{} survey. This includes their overall distributions on a wedge diagram, an evaluation of the quality of \hi{} data, and a comparative analysis of parameters with prior \hi{} surveys in the ZOA. Furthermore, we conclude with a search for multi-wavelength counterparts and an examination of the \hi{} properties distribution in Vela$-$\hi{}.

\subsection{General redshift wedge}\label{sec5:wedge}

    \begin{figure*}
	\centering
    \subfigure{\includegraphics[width=0.425\linewidth]{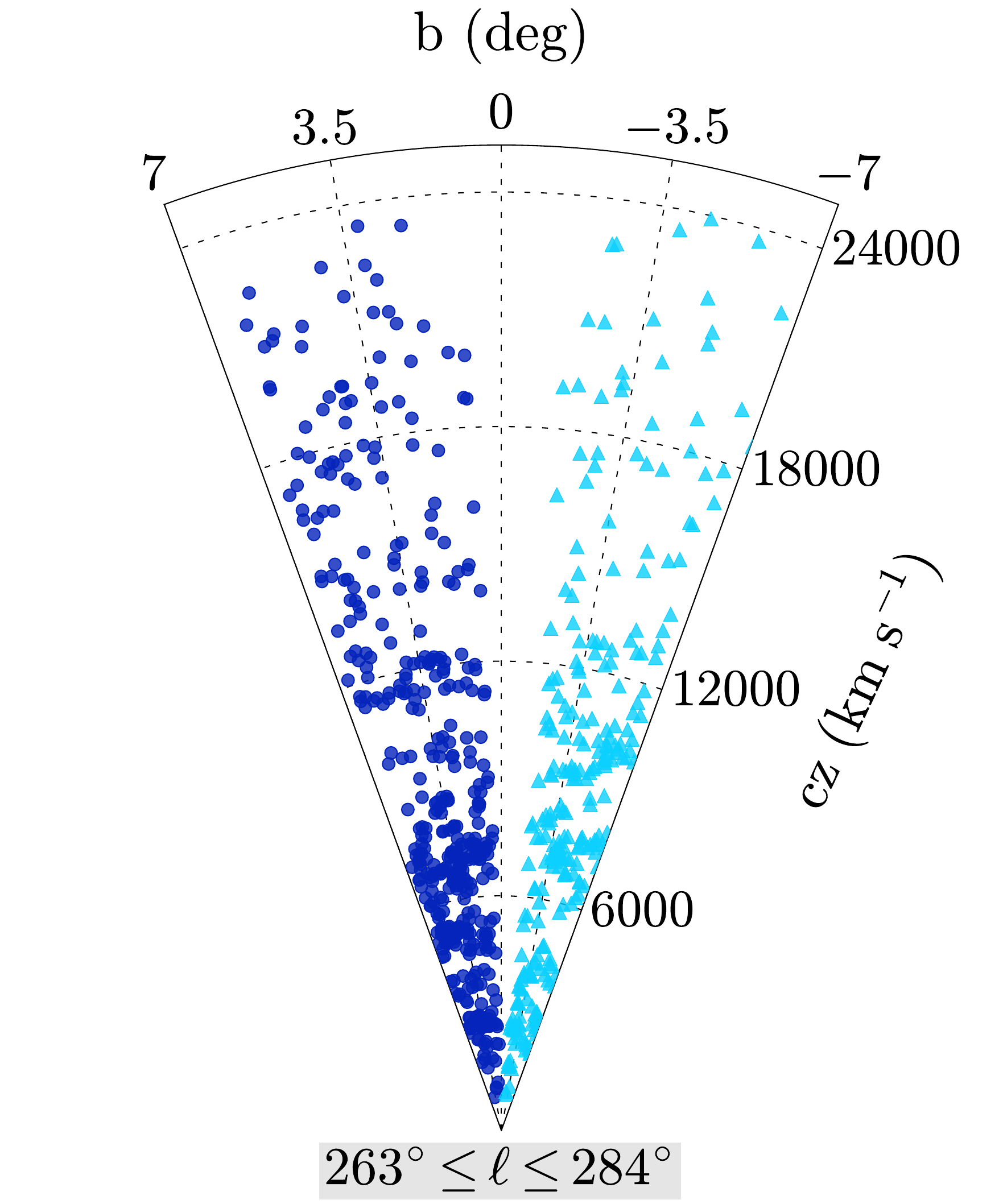}}
    \subfigure{\includegraphics[width=0.55\linewidth]{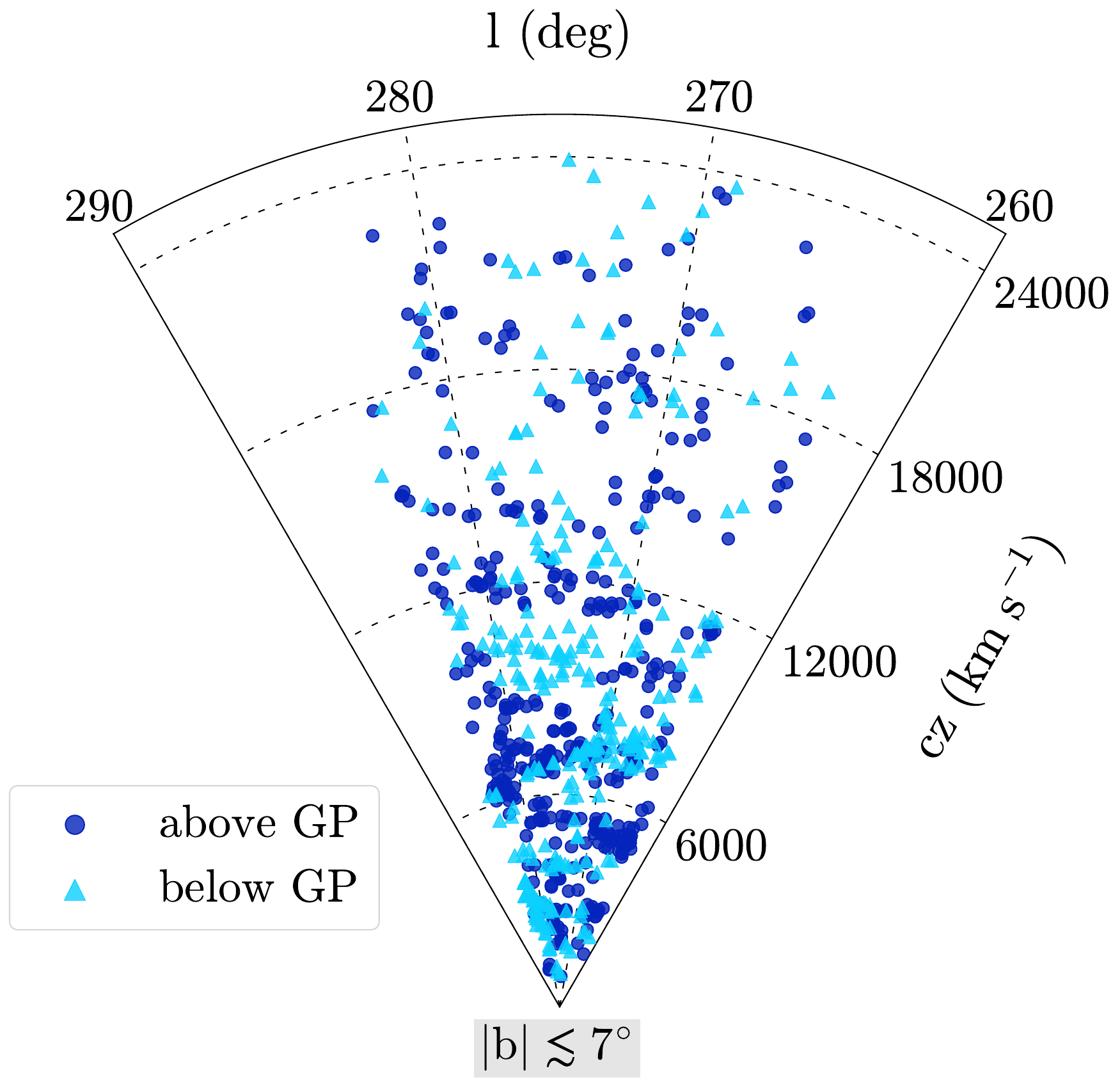}}
	\centering
	\caption{Latitude and longitude redshift wedge plots of detected galaxies in Vela$-$\hi\, represented by dark blue dots for those above the GP and cyan triangles for those below.}
	\label{fig:wedge_VSCL}
\end{figure*}

Among the 719 detections, 432 galaxies are detected above the GP, while 287 are situated below. Their distribution in latitude (left) and longitude (right) redshift slice diagrams across the full velocity range of the Vela$-$\hi{} ($cz < 24000$ \kms{}) is illustrated in Fig. \ref{fig:wedge_VSCL} where dark blue points indicate those above the GP and cyan triangles those below the GP. There are clear overdensities and distinct structures apparent in both panels. 
For instance, in the longitude wedge diagram, some structures, are prominent only below the GP (e.g., cyan triangles at ${\sim} 10000$ \kms{}), while others overlap, suggestive of potential connections across the ZOA (e.g., at ${\sim} 8000$ \kms{}). Further details regarding the LSS observed in the wedges will be discussed in Section \ref{sec5:LSS}.

\subsection{Internal quality assessment}\label{sec5:dupl_analysis}

Thanks to the presence of overlapping regions within the mosaics described in Section \ref{sec5:mosaicking}, we conducted an independent assessment of the \hi{} data quality. A total of 163 detections were found in one or more adjacent mosaics. 
The heliocentric velocities and 50\% linewidths of these sources exhibited only small median differences of $6.5 \pm 9.9$ \kms{} and $11.8 \pm 30.2$ \kms{}, respectively, measuring less than half a channel width ($\Delta v = 44.3$ \kms{} at $z = 0$). The median integrated flux difference was found to be 0.3 $\pm$ 0.8 Jy \kms{}, lower in most cases than the integrated flux errors of the individual detections. The high level of consistency in the \hi{} parameters from overlapping regions confirms the robustness of the derivation of these properties.

\begin{figure}
	\centering
	\includegraphics[width=0.8\linewidth]{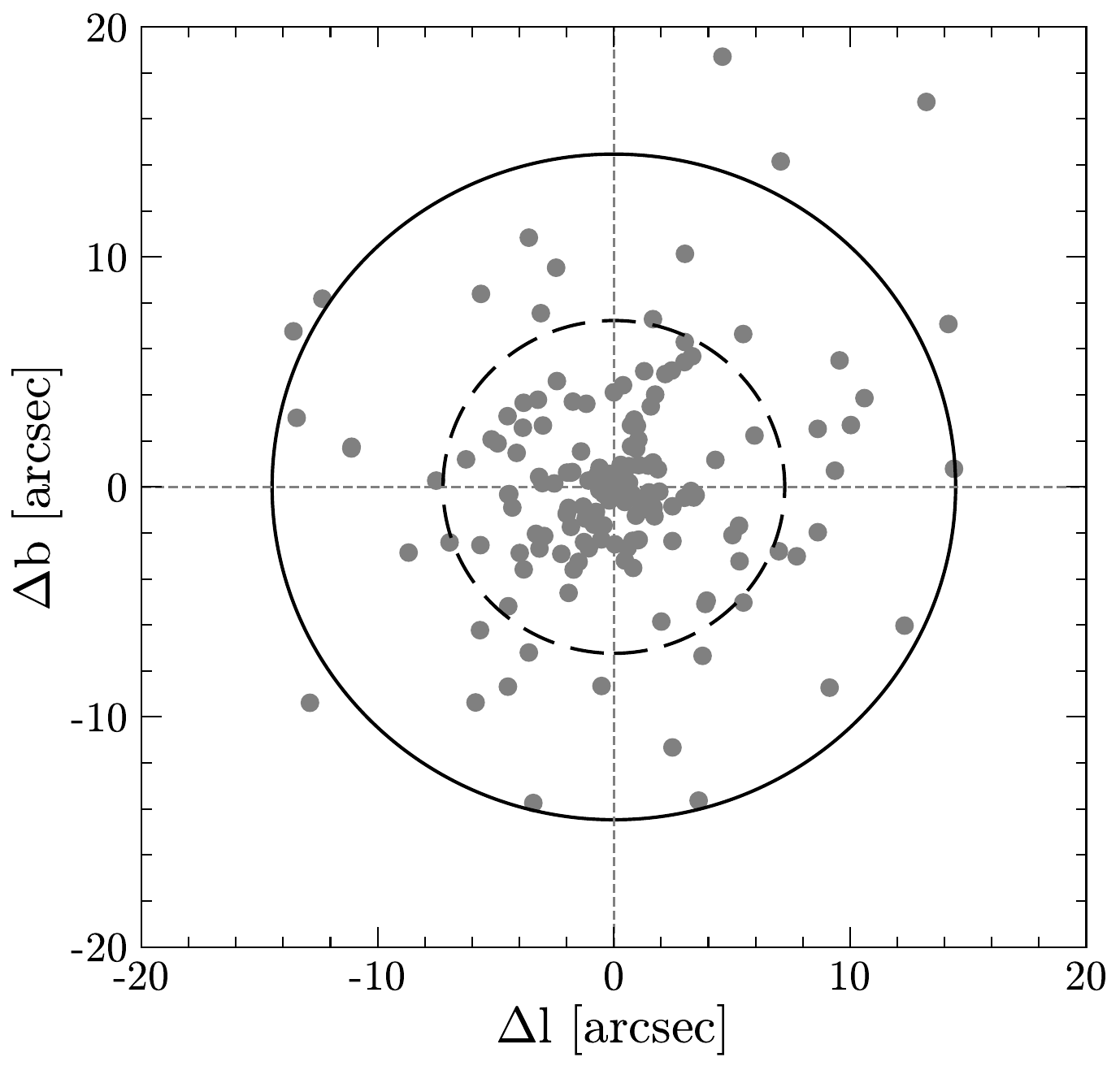}
	\centering
	\caption{Galactic coordinates separations between Vela$-$\hi{} detections found in the overlapping regions of the mosaics. The black dashed and solid circles mark the $1\sigma$ and $2\sigma$ of the coordinate separations.}
	\label{fig:coord_prec}
\end{figure}

We also used the duplicates to assess the positional accuracy of the survey. In Fig. \ref{fig:coord_prec}, the respective offsets are displayed. By measuring the differences in Galactic coordinates of each duplicate, we achieved an average coordinate precision of $\ang{;;6.04}$, with a $1\sigma$ standard deviation of $\ang{;;7.23}$. This value is close to three times larger than that of Vela$-$SMGPS, which can be attributed to the larger beam size of Vela$-$\hi{}. We did not distinguish between solid and possible detections as only two of the duplicates are classified as possible detections. The estimate of positional accuracy is an important parameter in our search for multi-wavelength counterparts (see Section \ref{sec5:multi_counterparts}).

\subsection{Comparison with other \hi{} surveys}\label{sec5:comparison}

We now explore any potential systematics in the measured \hi{} parameters of our survey compared to other \hi{} surveys that have some overlapping footprint with Vela$-$\hi{}, such as HIZOA and the Vela$-$SMGPS.

\subsubsection{Comparison with Vela$-$SMGPS}\label{sec5:comp_GPS}

\begin{figure*}
	\centering
	\subfigure{\includegraphics[width=0.48\linewidth]{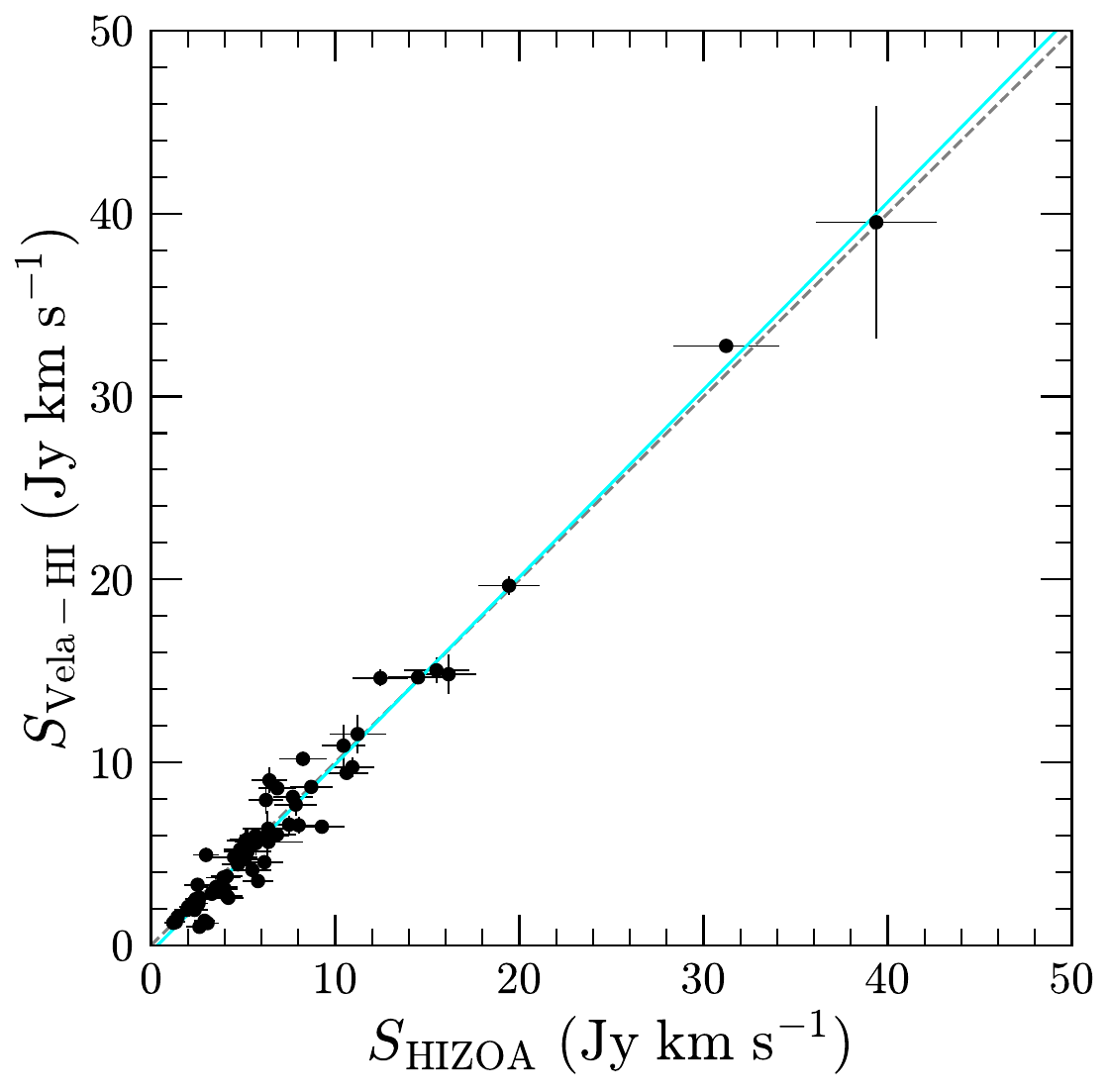}}
	\subfigure{\includegraphics[width=0.47\linewidth]{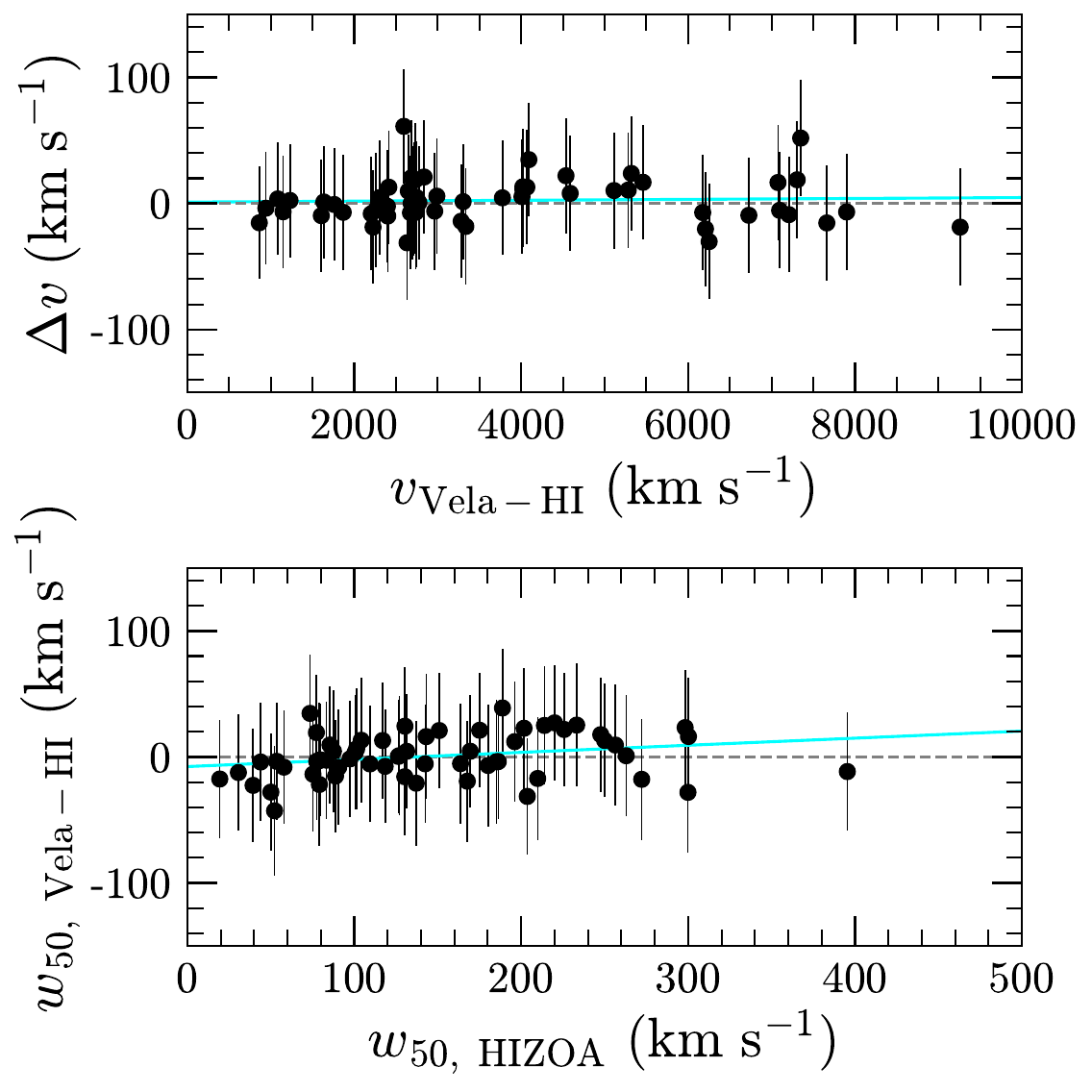}}
	\centering
	\caption{In the left panel, the integrated fluxes of Vela$-$\hi\ detections with known counterparts from HIZOA are compared. The one-to-one relation is represented by the black dashed line, and the best linear fit is indicated by the cyan solid line. In the right panels, the differences in systemic velocities (top panel) and \hi{} linewidth (bottom panel) between Vela$-$\hi\ and HIZOA are illustrated. The cyan lines in both panels depict the outcomes of the best-fit linear regressions, with the dashed black zero-difference line serving as a reference for comparison.}
	\label{fig:HIZOA_comp}
\end{figure*}

In comparing Vela$-$\hi{} with Vela$-$SMGPS, we identified 15 galaxies in common that have separations below $\ang{;;10}$ and velocity offsets limited to 75 \kms{}. The mean separation of these counterparts was $\ang{;;5.4} \pm \ang{;;2.7}$, well within the coordinate precisions of both surveys. However, 3 of these detections were situated in strong continuum regions at the edge of Vela$-$\hi{}, and 6 others had very low integrated signal-to-noise ratio (SNR $<$ 6) in Vela$-$\hi, making their fluxes unreliable for comparison. We therefore limited our comparison to the remaining 6 galaxies with SNR $>$ 9.

Upon examining differences in velocities $cz$,
and linewidths $w_{50}$, we noted that Vela$-$\hi{} galaxies exhibit slightly higher velocities compared to those in Vela$-$SMGPS, showing an offset of $9.6 \pm 5.1$ \kms{} and $9.1 \pm 5.7$ \kms{} in linewidth, respectively. The systemic velocity difference slightly increases with redshift, while the linewidth difference is smaller for larger linewidths, except for cases with low SNR and large $w_{50}$, where the emission peaks are easily hidden by noise. Despite these variations, all values fall within the one-channel resolution of both surveys and lie within the expected error margin.

In terms of fluxes, the linear regression fit found for the 6 galaxies in common follows a near one-to-one relation with a slope of $0.982 \pm 0.298$, an intercept of $-0.886  \pm 1.938$, and an R$^2$ value of 0.93. No statistically significant ($3\sigma$) systematic effects are noticeable for this small sample of galaxies with high SNR.\\

To further assess the reliability and completeness of our survey, we investigated the Vela$-$\hi{} detections not found in Vela$-$SMGPS, and vice versa. The three Vela$-$\hi{} galaxies not cataloged in Vela$-$SMGPS had a low SNR $< 5$ with a Flag of 2 (i.e., they were designated as possible detections). They were all positioned at the noisy edge of the Vela$-$SMGPS mosaics which could have led to them being missed by SoFiA. 

Conversely, 49 Vela$-$SMGPS galaxies (28 above and 21 below the GP) were not detected by Vela$-$\hi{}. This is not surprising given the higher sensitivity of Vela$-$SMGPS, with a mean rms of 0.39 \mJy{} compared to 0.74 \mJy{}. Galaxies undetected in Vela$-$\hi{} revealed a median SNR of 8, median integrated fluxes of 0.45 Jy \kms{}, median velocities around 11500 \kms{}, and a median $w_{50}$ of 145 \kms{} in Vela$-$SMGPS. Those with SNR $>$ 8 but undetected in Vela$-$\hi{} typically had local rms values below the average noise, indicating they were close to the detection limit threshold and located in low rms areas of Vela$-$SMGPS, hence too faint for our survey. Only galaxies with high SNR and relatively nearby were therefore likely to be detected in common by both surveys.

\subsubsection{Comparison with HIZOA}\label{sec5:comp_HIZOA}

We identified 87 HIZOA detections out of the 90 present within our surveyed area. Three sources -- J0932-44, J0916-55, and J0933-58 -- remained undetected. The absence of J0932-44 and J0916-55 in our Vela$-$\hi{} data can be attributed to their narrow $w_{50}$ linewidths in HIZOA (87 and 140 \kms{}, respectively). No $w_{20}$ information is also available in the HIZOA catalog, which could be part of the reason why they were overlooked given our coarse resolution. As for J0933-58, it was not recovered by SoFiA, likely due to (i) its location in an area strongly affected by residual solar RFI and (ii) its proximity to the edge of the source-finding region. Among the 87 HIZOA detections, 5 were resolved into multiple counterparts in the Vela$-$\hi{} survey.
These sources resolved into several counterparts were excluded from the comparison. To ensure a robust comparison, we also excluded sources with a low SNR (SNR $< 10$) in Vela$-$\hi{}, and those affected by residual solar RFI, leading to poor baselines. This filtering process led to 61 galaxies for comparison.

The remaining detections displayed a mean coordinate precision of $\ang{;1.9} \pm \ang{;1.6}$. The flux fit regression, illustrated in Fig. \ref{fig:HIZOA_comp}, reveals a slope close to unity ($1.024 \pm 0.042$), an intercept of $-0.358 \pm 0.405$, with R$^2 = 0.98$. The right panels of Fig. \ref{fig:HIZOA_comp} display the differences in systemic velocities and linewidths. The systemic velocity plot (top right panel) yields a mean absolute difference of $\Delta v \sim 12.2$ \kms{} with a standard deviation of 11.1 \kms{}, which is insignificant given the errors. The linewidth difference (bottom right panel) has a slight offset of $13.2 \pm 10.3$ \kms{}.
Given that linewidths have been corrected for instrumental resolution, and the coarse resolution (${\sim} 44$ \kms{}), neither this offset nor the trend (cyan line) is significant. The linewidths compared to HIZOA are consistent with each other; on a one-to-one basis, we find a slope close to unity ($1.009 \pm 0.058$) and an intercept of $-0.836 \pm 9.682$, with R$^2 = 0.95$.

\subsection{Multi-wavelength counterparts}\label{sec5:multi_counterparts}

The Vela$-$\hi{} survey uniquely connects the Vela$-$SMGPS to the higher latitudes where deep optical and NIR surveys like Vela$-$OPT/NIR, 2MRS, etc., become successful in unveiling LSS. Compared to SMGPS, this increases the likelihood of finding multi-wavelength counterparts.

We initiated a comprehensive search for multi-wavelength counterparts for our Vela$-$\hi{} detections using online databases such as the NASA/IPAC Extragalactic Database (NED)\footnote{http://ned.ipac.caltech.edu} and the Vizier 
Catalogue access tool \citep{Ochsenbein2000}. Data retrieval (e.g., tables) employed remote web querying forms facilitated by the Python module \texttt{Astroquery} \citep{Ginsburg2019}. 
The search radii were determined by the quadrature of the positional accuracy of the two cross-matched surveys. For instance, the search radius for \hi{} surveys like HIPASS \citep{Meyer2004} and HIZOA \citep{Staveley2016} typically spans a few arcminutes, while OPT and NIR catalogs have accuracies of a few arcseconds (e.g., 2MASX, has a positional accuracy of approximately $\ang{;;1}$, as detailed in \citealt{Jarrett2000}). There can also be offsets between the center of stellar and neutral hydrogen structures of a galaxy. We therefore applied a search radius of $\ang{;;15}$, considering the positional accuracy of Vela$-$\hi{} of ${\sim}\ang{;;6}$, and took into account sources classified as `G' or Galaxy in NED for star-galaxy separation. This approach resulted in a list of potential counterparts and redshifts where available. To assist in identifying the most likely counterpart, an examination of a $\ang{;2} \times \ang{;2}$ image stamp from the Digitized Sky Survey (DSS), 2MASS-K band, and WISE-1 band ($3.4 \, \rm \mu m$) was examined (using \textsc{Skyview}, 
a package in \textsc{Astroquery}).\\

\begin{figure}
	\centering
        \includegraphics[width=\linewidth]{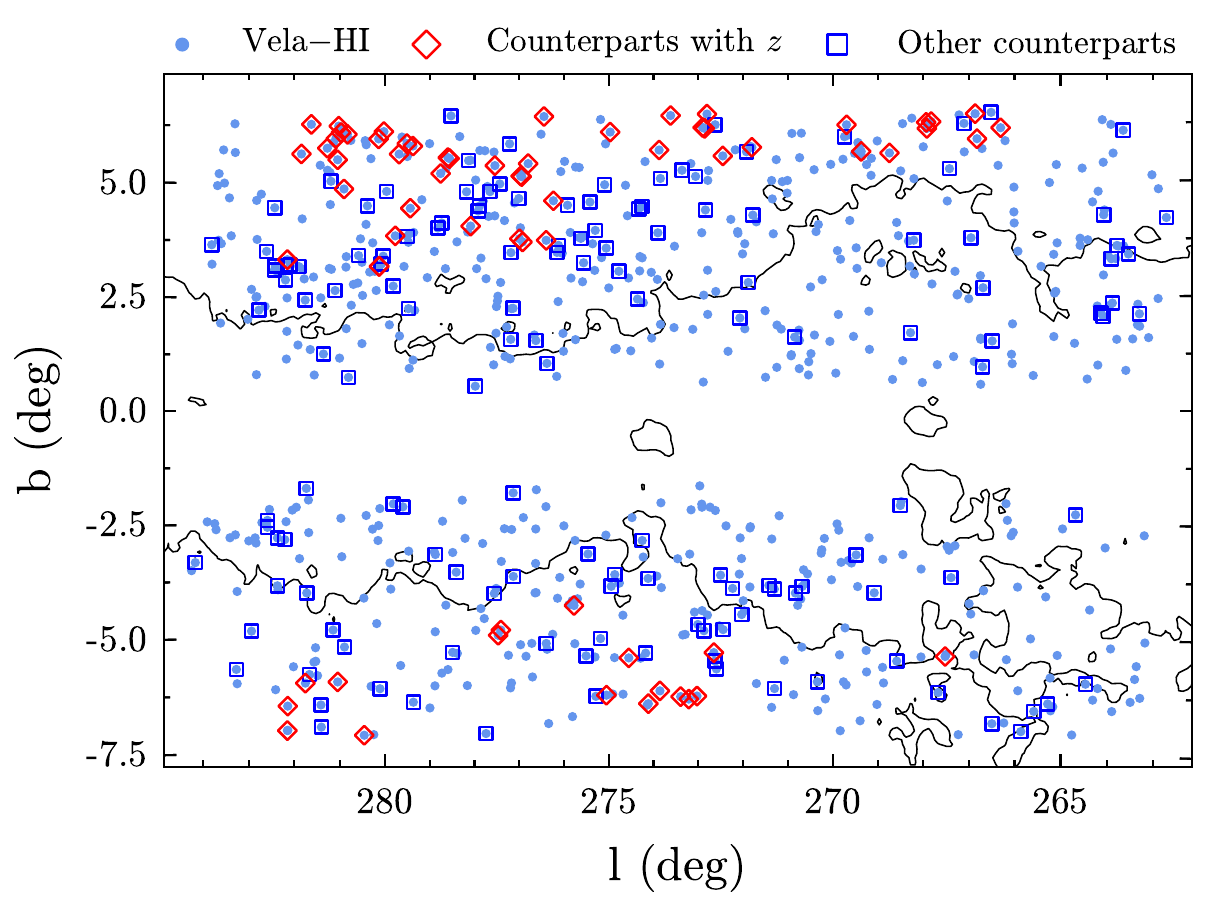}
	\centering
	\caption{On-sky distribution of Vela$-$\hi{} detections (light blue dots). Red open diamond symbols represent multi-wavelength counterparts with spectroscopic redshifts, while blue open
    squares have no redshift information. Black contour lines delimit the Galactic foreground extinction at
    $A_B \simeq 3.0 ~ \rm mag$.}
	\label{fig:counterpart}
\end{figure}

Figure \ref{fig:counterpart} illustrates the distribution of Vela$-$\hi{} detections and their potential multi-wavelength counterparts. Those with available optical spectroscopic redshifts are marked as red diamonds and those without redshift data with blue open squares. Below $|b| \lesssim 5^{\circ}$, the availability of redshift information diminishes due to increasing foreground extinction. This is shown by the extinction contour in grey representing the dust extinction limit in the optical of $A_B \simeq 3.0$ mag, derived from the Diffuse Infrared Background Experiment (DIRBE) maps \citep{Schlegel1998,Schlafly2011,Schroder2021}. Out of the 719 \hi{} detections, we identified 211 (${\sim}29$\%) probable counterparts. Their mean separation is minimal, measuring only $\ang{;;5.6} \pm \ang{;;3.4}$, less than the positional accuracy of Vela$-$\hi{}. 110 galaxies (${\sim}52$\% of 211) have redshift information, with 66 (${\sim}31$\% of 211) originating from optical spectroscopy, and 27 (${\sim}13$\% of 211) having \hi{} redshifts only (HIZOA and HIPASS). Photometric redshifts were not used for comparison given their huge uncertainties. 
For Vela$-$\hi{} galaxies with velocities exceeding 16000 \kms{}, i.e., at the VSCL distance, 19 have likely counterparts, but only 5 have available spectroscopic redshift data.\\

\begin{figure}
	\centering
	\includegraphics[width=\linewidth]{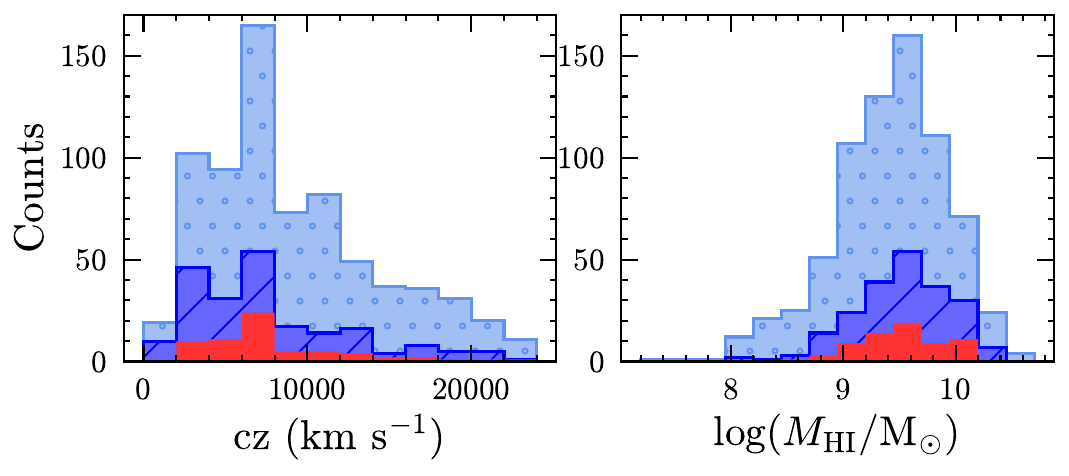}
	\centering
	\caption{Histograms comparing Vela$-$\hi{} detections (light-blue with dotted patterns) and their likely counterparts as a function of velocity (left panel) and \hi{} mass (right panel). Galaxies with counterparts are depicted by the blue-hatched histograms, and counterparts with spectroscopic redshifts are represented by red-shaded histograms.}
	\label{fig:counterpart_hist}
\end{figure}

The histograms in Fig. \ref{fig:counterpart_hist} compare the distribution in velocity and \hi{} mass for the Vela$-$\hi{} detections (light-blue with dotted patterns) with their likely counterparts (hatched blue). Counterparts with spectroscopic redshift information are depicted in red. In the left panels, we observe a decrease in the number of counterparts as velocities increase. The peaks in counterpart distributions align with overdensities in Vela$-$\hi{} (see Section \ref{sec5:sens_curve}). In the right panel, galaxies with low \hi{} masses $\log (M_{\mathrm{HI}}/$ $\rm M_{\odot}) < 8.4$ have few matches. They are likely blue low surface brightness dwarfs, therefore difficult to see through extinction layers.
There is a slight increase in the number of cross-identified \hi{} detections with increasing \hi{} masses, which are generally large spirals that are prominent in optical as well. They peak at the same point as the \hi{} mass distributions of Vela$-$\hi{}. The results from Figs. \ref{fig:counterpart} and \ref{fig:counterpart_hist} suggest that the likelihood of an \hi{} source having a potential counterpart increases with \hi{} mass but decreases as foreground extinction rises at lower latitudes. Moreover, there are no systematic surveys in the ZOA that cover higher redshifts. The only systematic survey is 2MASS, but it is also incomplete with increasing redshift, brightness-limited, and affected mostly by stellar crowding at lower latitudes, making massive and nearby galaxies likely to have more counterparts than dwarfs and high-$z$ galaxies.

\begin{figure}
	\centering
        \includegraphics[width=\linewidth]{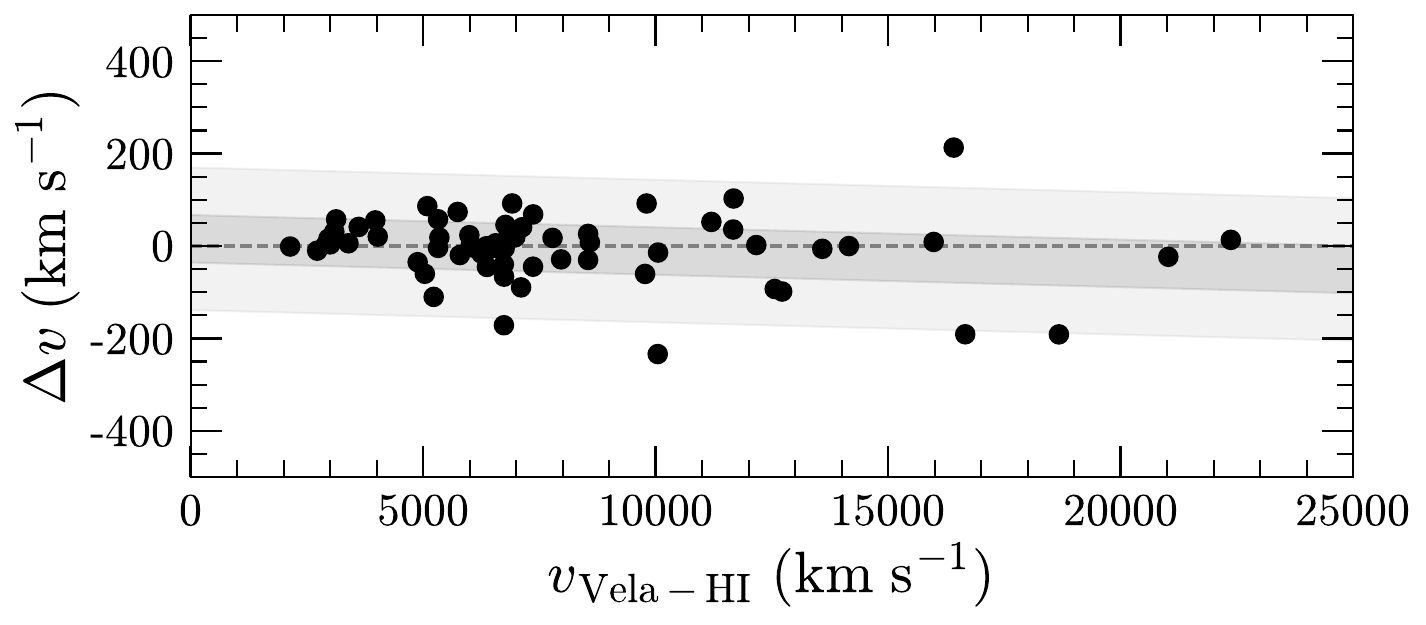}
	\centering
	\caption{Heliocentric optical velocity comparison between Vela$-$\hi{} galaxies and counterparts with spectroscopic redshifts ($\Delta v = v_{\rm Vela-HI} - v_{\rm OPT/NIR}$).
    The dark-shaded regions delineate the $1\sigma$ and $3\sigma$ standard deviations, respectively.
    }
	\label{fig:vopt_diff}
\end{figure}

We will now investigate the correspondence between spectroscopic and \hi{} redshifts. A one-to-one comparison results in a slope close to unity ($0.997 \pm 0.004$) with an R$^2 = 0.999$ and an intercept of $17.7 \pm 36.2$. These findings demonstrate a high level of consistency. A more detailed analysis, examining the difference between \hi{} and optical velocities ($\Delta v = v_{\rm Vela-HI} - v_{\rm opt}$) as illustrated in Fig. \ref{fig:vopt_diff} reveals an average difference of $70 \pm 51$ \kms{}.
The best-fitting parameters of the velocity differences have a slope and intercept of $-0.003 \pm 0.004$ and $15.6 \pm 36.4$, respectively, showing no significant difference with increasing velocities. In Section \ref{sec5:comp_HIZOA}, the comparison with HIZOA of the Vela$-$\hi{} velocities found uncertainties to be of the order of $13.2 \pm 10.3$ \kms{}. Errors compared to the optical velocities, generated from multi-object spectroscopy such as AAOmega, have uncertainties on the order of ${\sim}100$ \kms{} \citep{Baldry2014}. The determined offsets between optical and Vela$-$\hi{} velocities range from 0.14 \kms{} to 234 \kms{}.\\

In summary, of the 211 identified galaxies, 146 (70\%) are associated with 2MASS/2MASX counterparts \citep{Jarrett2000, Skrutskie2006}, 140 (66\%) have WISE counterparts \citep{Wright2010, Cutri2014}, 92 (44\%) are associated with HIZOA detections \citep{Staveley2016}, 49 (23\%) have IRSF counterparts \citep{Williams2014,Said2016}, 38 (18\%) have IRAS counterparts \citep{Helou1988}, and 35 (17\%) are linked to HIPASS counterparts \citep{Meyer2004}. Of the 92 Vela$-$\hi{} detections with likely HIZOA counterparts, 27 have probable WISE counterparts. A noteworthy aspect of Vela$-$\hi{} is its partial overlap with the Vela$-$OPT/NIR survey. More than half (121 out of 211) of the identified counterparts originate from the overlap between Vela$-$\hi{} and Vela$-$OPT/NIR. These are labeled with prefixes `HyA' and `Vel' in the last column of the galaxy catalog (see online supplementary material).

\subsection{\HI{} properties distributions}\label{sec5:hi_distribution}

\begin{figure}
	\centering
	\includegraphics[width=\linewidth]{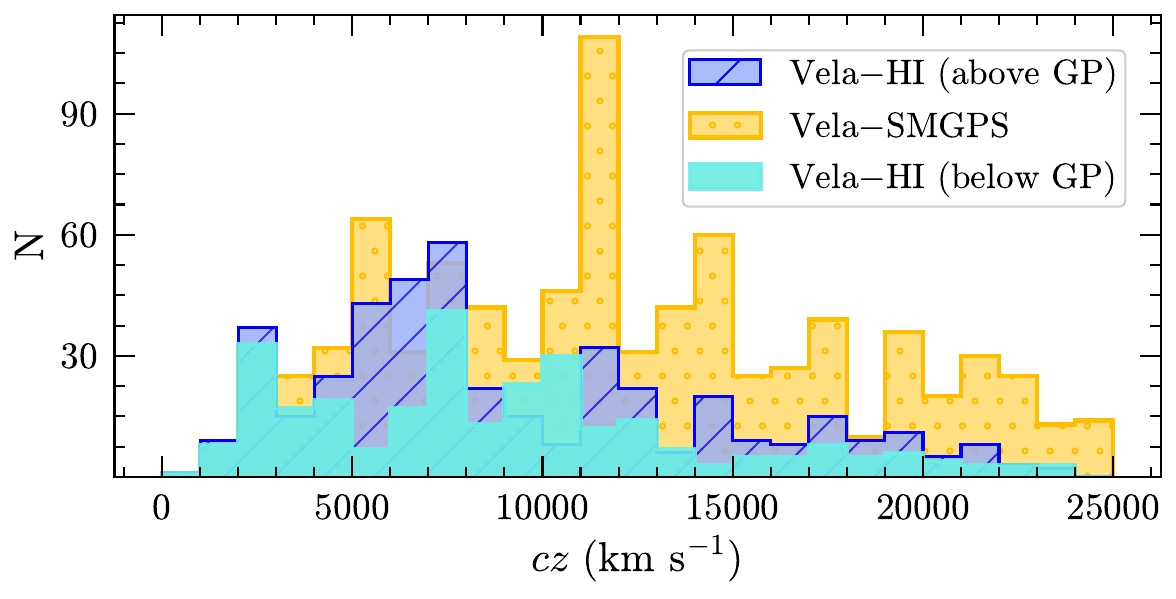}
	\caption{Histograms showing the heliocentric velocity distributions of the galaxies in Vela$-$\hi{} above the Plane (hatched blue), Vela$-$SMGPS (orange with dotted patterns), and Vela$-$\hi\ galaxies below the GP (cyan), respectively.}
	\label{fig:hist_vel}
\end{figure}

\begin{figure}
	\centering
	\includegraphics[width=\linewidth]{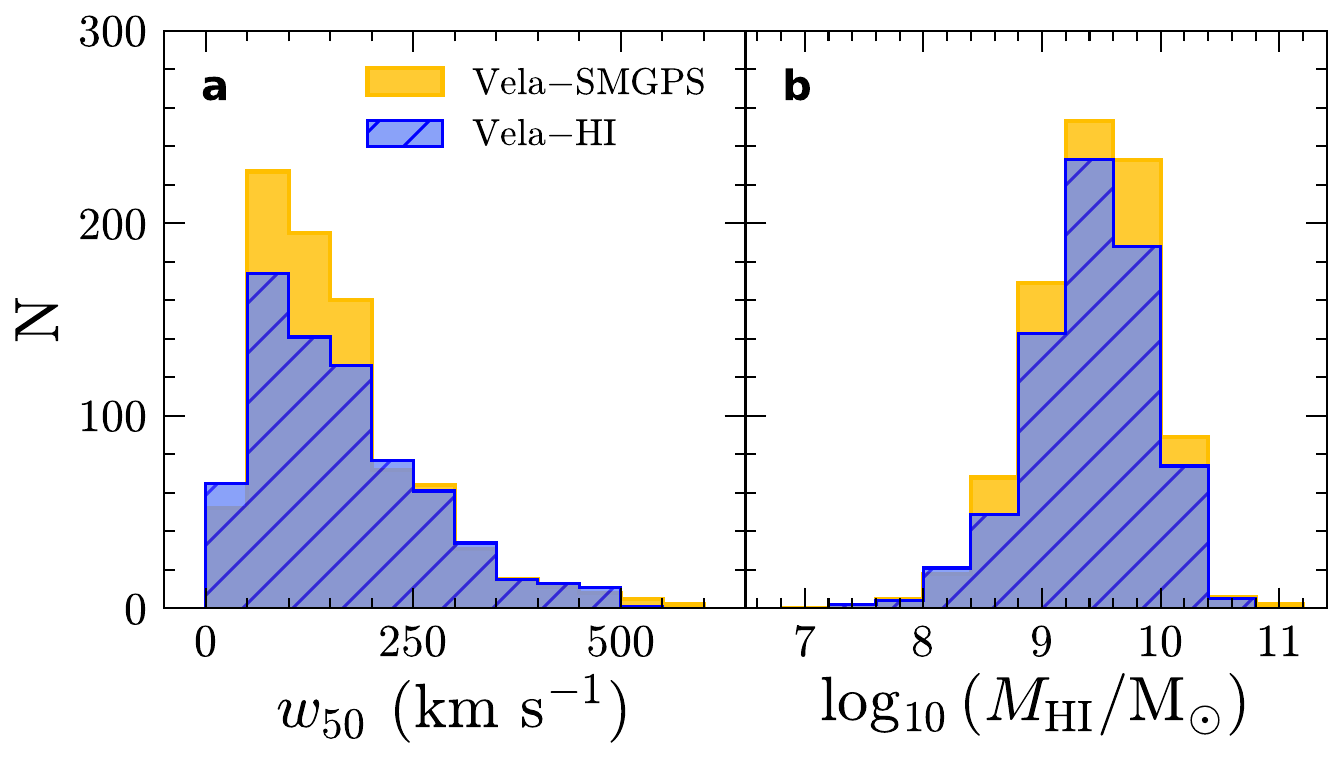}
	\centering
	\caption{Histograms of the \hi\ parameters from the systematic blind \hi\ surveys Vela$-$\hi\ (719 galaxies, hatched blue) and Vela$-$SMGPS (843 galaxies, orange). Left panel (a) displays the linewidths ($w_{50}$) and right panel (b) shows the \hi\ mass distribution of detected galaxies.}
	\label{fig:hist_VSCL_GPS}
\end{figure}

As a next step, we explore the overall distribution of the global \hi{} properties between Vela$-$\hi{} and Vela$-$SMGPS such as the recessional velocities (Fig. \ref{fig:hist_vel}), the linewidths ($w_{50}$) and \hi{} masses (Fig. \ref{fig:hist_VSCL_GPS}).\\ 

The heliocentric velocity distributions of Vela$-$\hi{} and Vela$-$SMGPS are shown in Fig. \ref{fig:hist_vel}. The hatched blue histogram indicates galaxies above the GP, the orange histogram with the dotted pattern shows galaxies in Vela$-$SMGPS, and the cyan histogram shows galaxies below the GP.  Because of the lower sensitivity of Vela$-$\hi{}, the median velocity is lower (7700 \kms{}) compared to that of Vela$-$SMGPS (${\sim}11600$ \kms{}). However, both surveys uncover galaxies out to the redshift limit of the surveys (24000 \kms{}). Several prominent peaks are discernible across regions above, within, and below the GP, suggesting a continuity of LSS. While the significance of these peaks will be discussed in detail in Section \ref{sec5:LSS}, we identify and highlight some of them here.

Four peaks stand out in Fig. \ref{fig:hist_vel} at $2000-3000$ \kms{}, $5000-8000$ \kms{}, $11000-12000$ \kms{}, and $14000-15000$ \kms{}. The first peak is observed in both surveys. The second peak is broad, with its higher velocity range ($7000-8000$ \kms{}) present both above and below the GP in Vela$-$\hi{}. 
The third peak has its origin entirely from the narrow filament unexpectedly discovered at $11000-12000$ \kms{} in Vela$-$SMGPS. Finally, the fourth peak is prominently seen in Vela$-$SMGPS, with a hint also observed above the GP in Vela$-$\hi{}.\\

Despite Vela$-$\hi{} covering a larger area with an rms nearly twice that of Vela$-$SMGPS, the distributions of linewidth and \hi{} mass in both surveys (Fig. \ref{fig:hist_VSCL_GPS}) exhibit similar patterns. The average linewidths of 162 and 158 \kms{} are practically the same for Vela$-$\hi{} and Vela$-$SMGPS, respectively, given the coarse velocity channel width. Both show peaks between 50 and 100 \kms{} linewidth bin, with Vela$-$SMGPS showing more narrow linewidth galaxies between 50 to 200 \kms{}. They both follow the skewed pattern typical of blind systematic \hi{} surveys (cf. HIZOA, \citealt{Staveley2016}; the EBHIS Zone of Avoidance survey or EZOA, \citealt{Schroder2019}). This contrasts to optically or NIR-selected \hi{} surveys, which follow a more Gaussian distribution with a rise towards larger linewidths ($w_{50} > 200$ \kms{}) because these tend to miss the bluer, low surface brightness dwarf population, as exemplified by \cite{Kraan2018}.

The distribution of \hi{} masses (panel b) reveals about three orders of magnitudes in $\log (M_{\mathrm{HI}}/\rm M_{\odot})$, 7.3 to 10.6 in Vela$-$\hi{} and 7.8 to 10.9 in Vela$-$SMGPS. They both peak between $9.2-9.6$, and have the same mean at $\log (M_{\mathrm{HI}}/\rm M_{\odot}) = 9.4$, despite Vela$-$SMGPS having a slightly higher number of galaxies per bin.


\section{Large-scale structures}\label{sec5:LSS}

\begin{figure*}
	\centering
	\includegraphics[width=\linewidth]{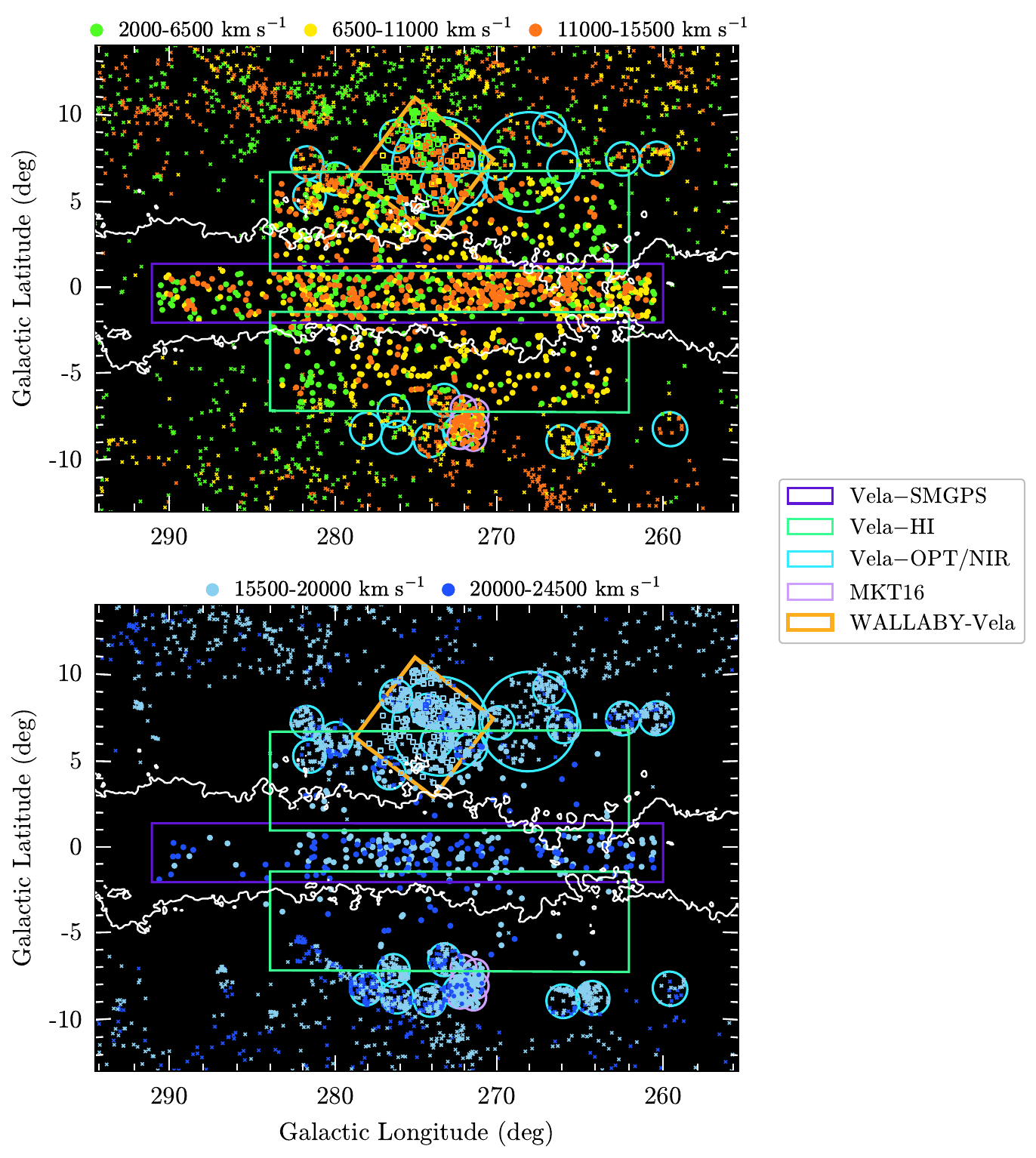}
	\centering
	\caption{Spatial distributions in Galactic coordinates of galaxies surveyed across the wider Vela region ($256^{\circ} < \ell < 294^{\circ}, |b| < 15^{\circ}$) with velocities reaching out to 24500 \kms{} from prior optical and \hi{} redshift surveys, combined with the systematic MeerKAT \hi{} surveys Vela$-$\hi{} and Vela$-$SMGPS. The top panel shows nearby velocity ranges at $2000-6500$ \kms{} (light green), $6500-11000$ \kms{} (yellow), and $11000-15500$ \kms{} (orange), while the bottom panel covers the higher velocity range near the VSCL distance: $15500-20000$ \kms{} (light blue), and $20000-24500$ \kms{} (dark blue). Large dots within the green and purple rectangles represent newly identified \hi{}-detected galaxies from Vela$-$\hi{} ($N=719$), and previously detected \hi{} galaxies from Vela$-$SMGPS ($N=843$) in the inner ZOA. Small crosses within cyan circles mark galaxies with redshifts obtained from Vela$-$OPT/NIR ($N=3013$), while small crosses outside the circles encompass all other compiled redshifts from various surveys (HyperLEDA and 2MRS). MeerKAT16 survey detections ($N=156$) appear as small dots within the six light purple circles around $b = -7^{\circ}$ \citep{Steyn2023_Msc}. WALLABY-Vela detections (\citealt{Murugeshan2024}, $N=143$) are represented by open square symbols within the orange diamond-shaped region. The white contour outlines corrected dust extinction at $A_B \simeq 3.0$ mag based on the DIRBE maps \citep{Schlegel1998,Schlafly2011}.}
	\label{fig:full_on_sky}
\end{figure*}

\subsection{Observational data in the Vela region}\label{subsec5:obs}

In this section, we discuss connections between previously identified and newly revealed LSS. As a starting point, we present Fig. \ref{fig:full_on_sky} to visually guide the interpretation of these structures. It combines MeerKAT data from heavily obscured galaxies acquired through Vela$-$SMGPS and Vela$-$\hi{} (purple and green boxes), along with ancillary datasets near the ZOA. These include galaxies with known redshifts from Vela$-$OPT/NIR (cyan circles), MeerKAT16 \hi{} observations of the surrounding of the galaxy cluster VC04 ($272.25^{\circ}, -9^{\circ}, 18000$ \kms{}, \citealt{Hatamkhani2023}), embedded in Wall 1 of the VSCL (\citealt{Steyn2023_Msc}; light purple circles). Those at higher latitudes ($5^{\circ} \leq |b| \leq 15^{\circ}$) have redshifts from the 2MASS Redshift Survey \citep{Huchra2012,Macri2019} and the HyperLEDA database \citep{Makarov2014}, represented by small crosses. Recent detections from the WALLABY-Vela field \citep{Murugeshan2024} are shown as open square symbols within the orange diamond shape above the GP. Different data point colors correspond to velocity ranges in intervals of 4500 \kms{}. The top panel displays three nearby velocity intervals, spanning from 2000 to 15500 \kms{}, which is the range probed by most surveys. The bottom panel presents the high-velocity range from 15500 to 24500 \kms{}, highlighting our new coverage with MeerKAT. The prominence of light and dark blue dots is already noticeable, with light blue indicating the distance range of Wall 1 and dark blue for Wall 2.

The region of high extinction, where optical and NIR galaxy counts become highly incomplete, is delineated by the white extinction contour. 
Caution is therefore advised when interpreting this on-sky plot and later LSS plots, as the presence of gaps around this contour does not necessarily indicate an area of underdensity.

The ZOA has now been quite thoroughly surveyed for the longitudinal range of $263^{\circ} < \ell < 284^{\circ}$. The new Vela$-$\hi{} detections, with its sensitivity of $0.74$ \mJy{} extend out to 24000 \kms. Due to the difference in rms, the coverage in Vela$-$\hi{} is approximately 3 galaxies per square degree compared to Vela$-$SMGPS with 9 galaxies per square degree. Both surveys highlight the effectiveness of systematic interferometric \hi{} surveys in mapping LSS across the most obscured part of the ZOA, representing a significant step toward understanding the continuity of LSS.

\subsection{Sensitivity curve and comparison with simulations}\label{sec5:sens_curve}

\begin{figure}
	\centering
	\includegraphics[width=\linewidth]{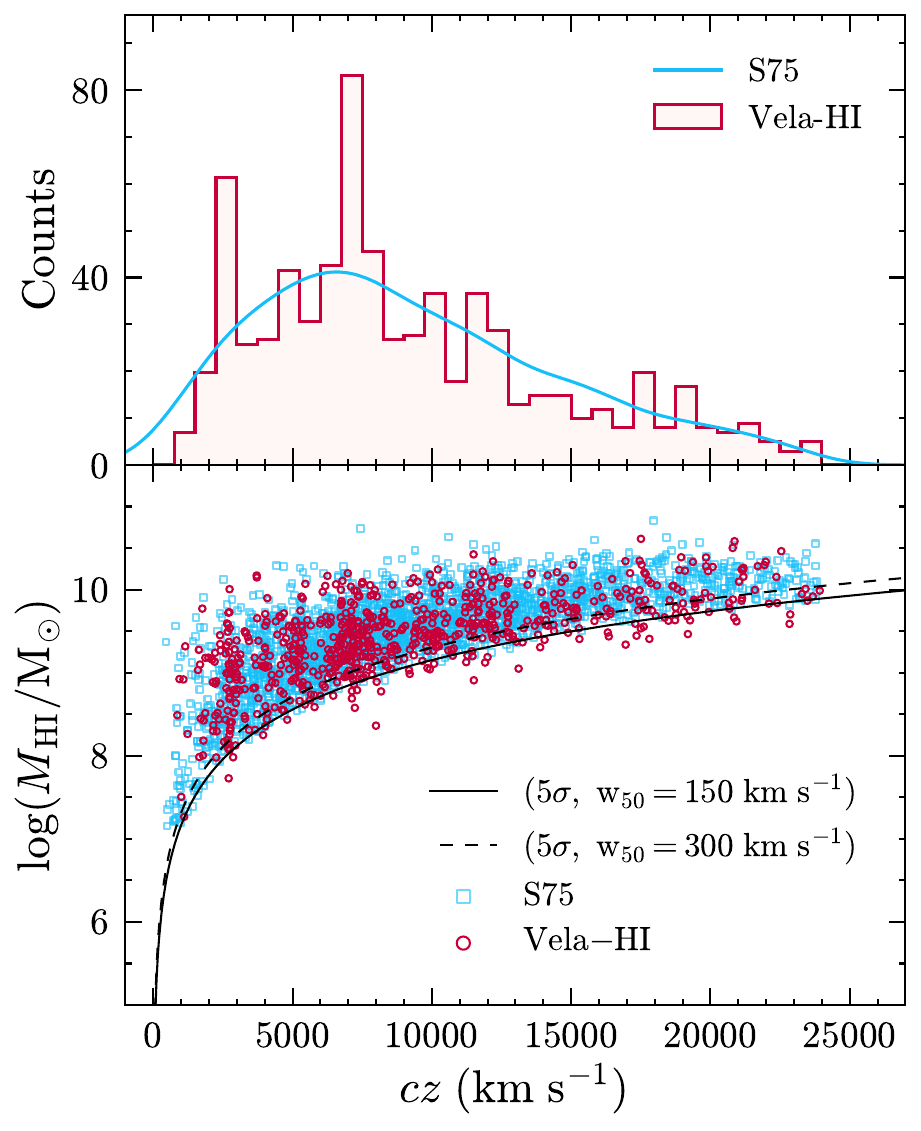}
	\centering
	\caption{The top panel presents the histogram distribution of heliocentric velocities in light red. The cyan solid line corresponds to simulated galaxies for the full SMGPS (528 deg$^2$) with an rms of 0.75 \mJy{} (S75) when scaled to the Vela$-$\hi{} survey area (193 deg$^2$). The bottom panel shows a sensitivity curve illustrating the correlation between \hi{} masses and heliocentric velocities for Vela$-$\hi{} detections (depicted as red dots), accompanied by the simulated galaxies in S75 (cyan squares). The solid and dashed lines represent the $5\sigma$ detection limits for $w_{50}$ values of 150 and 300 \kms{}, respectively.}
	\label{fig:sensitivity_curve_v2}
\end{figure}

As an initial step in understanding how the overdensities within the inner ZOA align with known LSS, we analyze the velocity distribution and sensitivity curve of Vela$-$\hi{} detections across the survey volume (see Fig. \ref{fig:sensitivity_curve_v2}). Similar to the simulation S45 presented in Vela$-$SMGPS (see \citealt{Rajohnson2024}), which assumes constant \hi{} mass and velocity width functions across redshifts to model a uniform galaxy distribution, we overlaid our \hi{} detections on a simulation that matches the sensitivity (0.75\,\mJy{}) of Vela$-$\hi{}. We focused on the survey area where it is complete (193\,deg$^2$) after removing the high-noise regions toward the outermost edges and the mosaic A2B region. This simulation, hereafter referred to as S75, predicts
769 galaxies for Vela$-$\hi{}. 

The prediction from the S75 velocity distribution, representative of a more uniform distribution, is shown by the cyan solid line. 
It reveals numerous over- and underdensities that are inconsistent with the uniformity (cyan line), clustering around ${\sim}2500$ \kms{}, $7000$ \kms{}, $17000-18000$ \kms{} and $19000-20000$ \kms{}. The last two peaks coincide with the redshift range of W1 and show an elongated distribution with \hi{} mass. A hint of a broader but more shallow overdensity at $21000-24000$ \kms{} is also noticeable. Towards the line-of-sight of VSCL, an underdensity from $13000-17000$ \kms{} is observed.

\subsection{Discussion of overdensities in the Vela region}\label{subsec5:lss}

In this section, we examine the on-sky distributions across varied redshift intervals, each for an interval of $\Delta v = 4500$ \kms{} increment range, with the aim of locating and identifying the distinct peaks more clearly. The redshifts in each 4500 \kms{} panel are subdivided into three increasing intervals of $\Delta v = 1500$ \kms{}, from cyan, green, to blue color symbols, respectively. 

\begin{figure*}
\centering
    \subfigure{\includegraphics[width=0.85\textwidth]{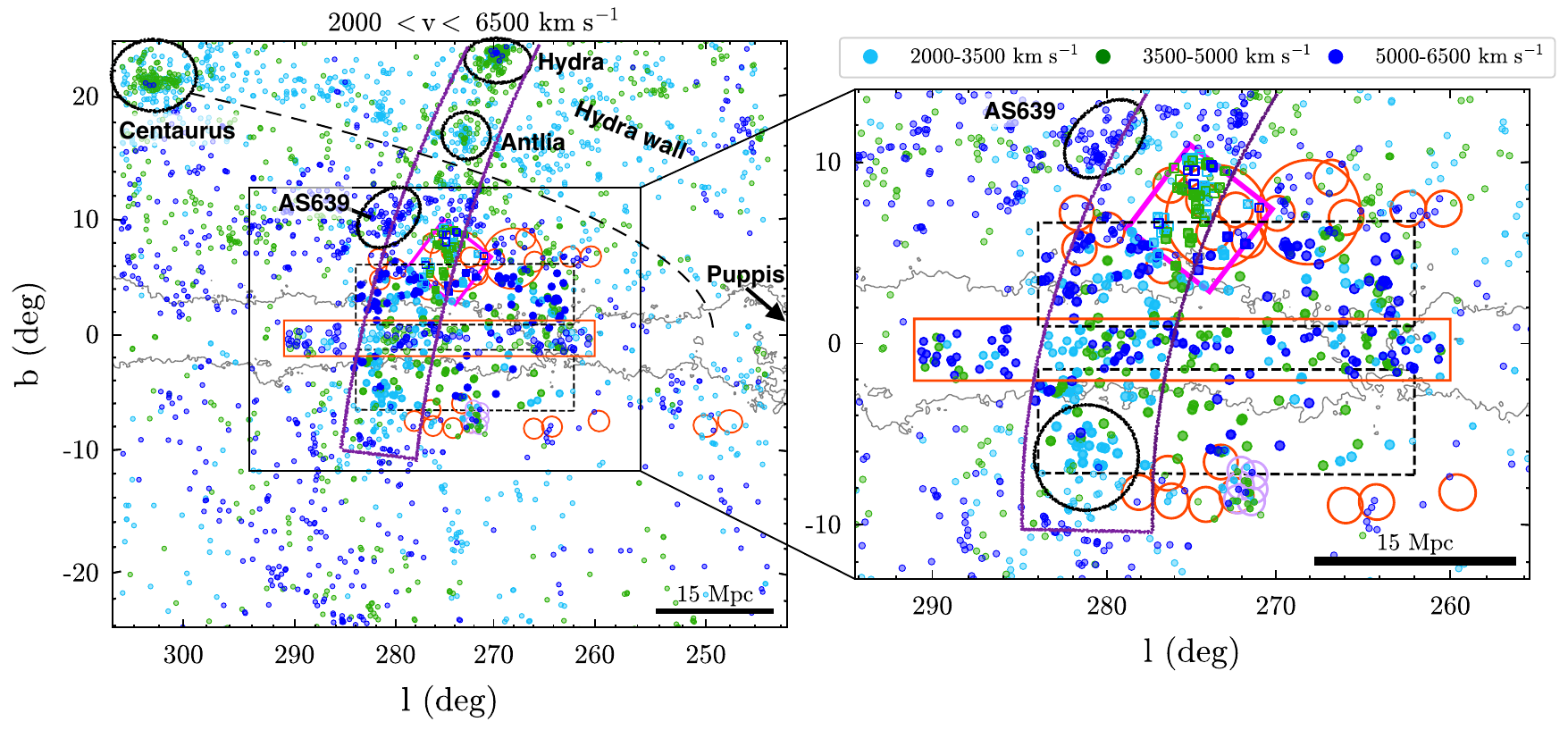}}\vspace{-1.5em}
    \hfill
    \subfigure{\includegraphics[width=0.85\textwidth]{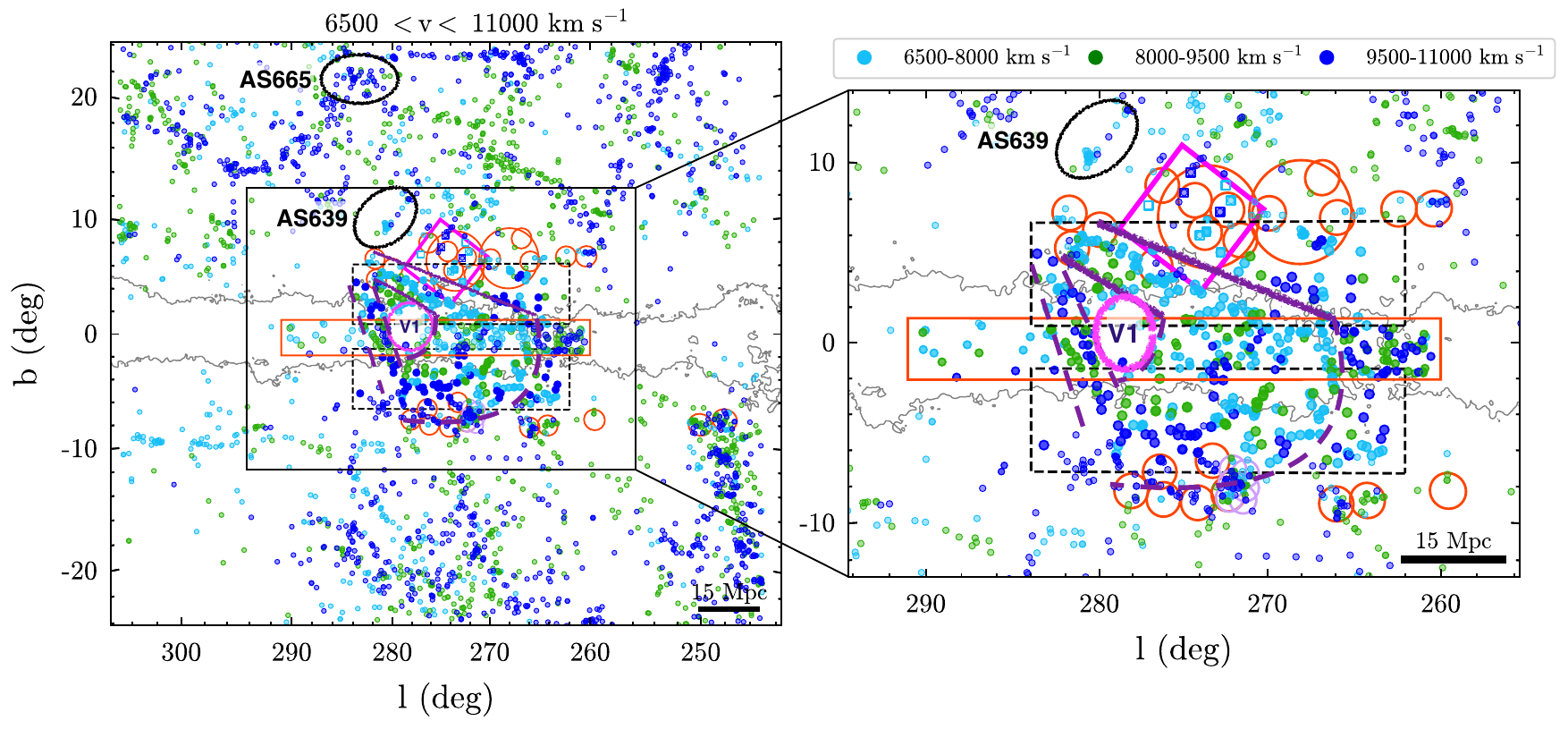}}\vspace{-1.5em}
    \hfill
    \subfigure{\includegraphics[width=0.85\textwidth]{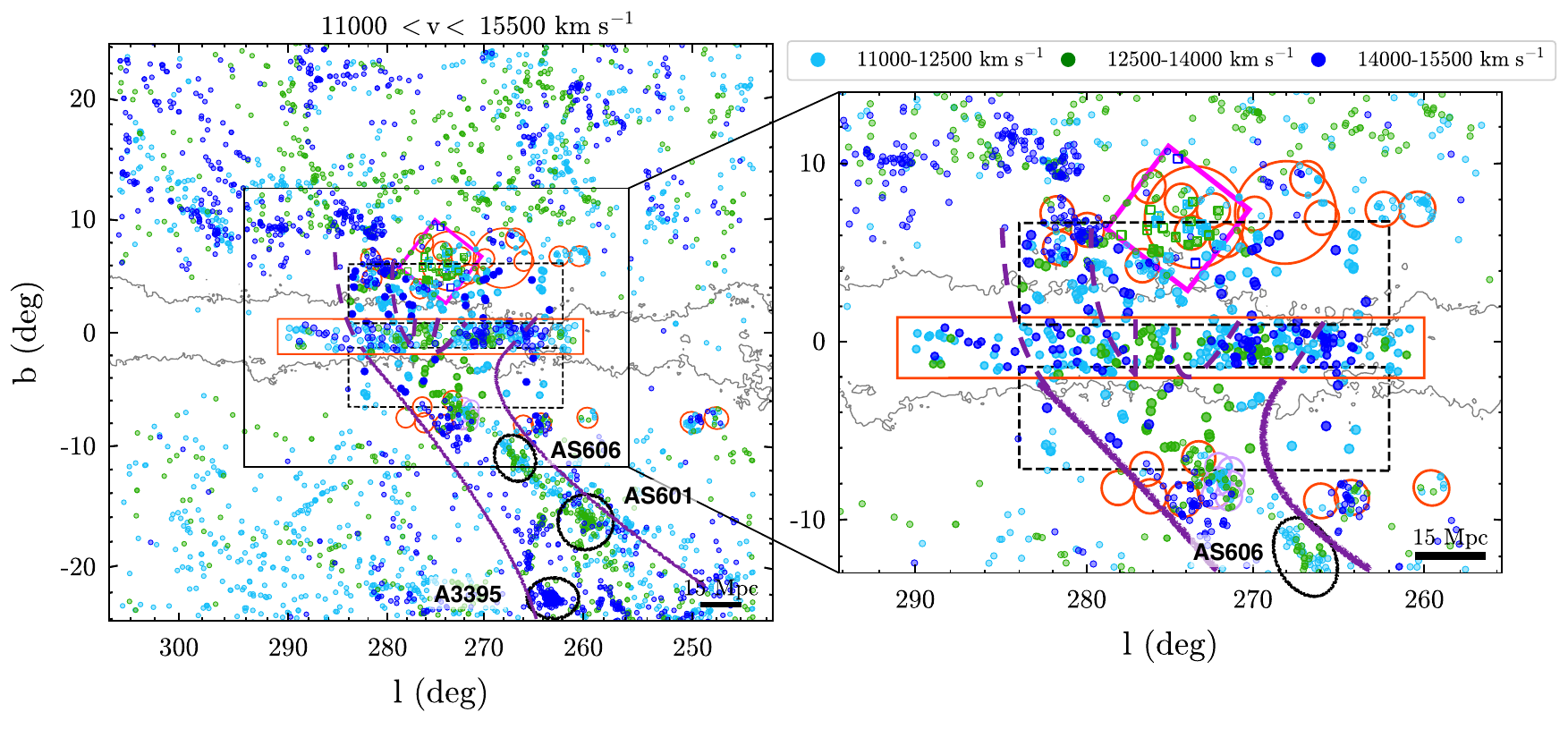}}\vspace{-1.5em}
	\caption{Annotated 2D on-sky distributions of galaxies in the Vela region and its surroundings ($244^{\circ} < \ell < 306^{\circ}, |b| < 25^{\circ})$. The figure displays three panels out to 15500 \kms{}, each representing a redshift interval with a $\Delta v = $ 4500 \kms{} from 2000 \kms{}. The color codes within each panel mark increasing velocity bins of $\Delta v = 1500$ \kms{}, from cyan, green, to blue. The newly discovered \hi{} galaxies in Vela$-$\hi{} are outlined by black dashed boxes, and Vela$-$SMGPS lie within the orange rectangle. Vela$-$OPT/NIR and MKT16 observations are denoted by orange and light purple open circles, respectively, while the WALLABY-Vela detections are represented by the open square symbols within the magenta diamond-shaped footprint. The dust extinction contour at $A_B \simeq 3.0$ mag is represented by a solid grey line. Each panel includes a zoomed-in section within $256^{\circ} \leq \ell \leq 294^{\circ}, |b| < 15^{\circ}$ and a 15 Mpc scale bar positioned at its lower right side.}
	\label{fig:LSS2_and_3D}
\end{figure*}

We first focus on nearby structures within the velocity range $2000 < cz \leq 15500$ \kms{} (see Fig. \ref{fig:LSS2_and_3D}), followed by galaxies at the distance range that encloses VSCL ($cz > 15500$ \kms{}; Fig. \ref{fig:LSS2_and_3D_2}). A zoom of the on-sky plots ($256^{\circ} \leq \ell \leq 294^{\circ}, |b| < 15^{\circ}$) is presented in the right-hand panel of Figs. \ref{fig:LSS2_and_3D} and \ref{fig:LSS2_and_3D_2}. We additionally present redshift wedges out to $cz < 25000$ \kms{} in Fig. \ref{fig:LSS_wedge_ann}, which include galaxies that lie beyond the ZOA. These data are represented by blue and orange dots for those above (left panel) and below (right panel) the GP, respectively.

\subsubsection{Nearby Structures: $cz \leq 15500$ \kms{}}\label{sec5:nearby_LSS}

$\mathit{2000} < cz \leq \mathit{6500} ~km\,s^{-1}$: A substantial filament extends from the Hydra ($270^{\circ}, +27^{\circ}$, 3400 \kms{}) and Antlia ($273^{\circ}, +19^{\circ}, 2600$ \kms{}) clusters, often referred to as the Hydra Wall across the GP \citep{kraan1994,kraan1995,Fairall1998,Kraan2000_vela,kraan2002}. Cyan and green points from the Vela$-$\hi{} together with the Vela$-$SMGPS data \citep{Rajohnson2024} trace the Hydra wall for the first time as a contiguous structure across the ZOA. The alignment of blue and orange points in the wedge slices (both panels in Fig. \ref{fig:LSS_wedge_ann}) further confirms these results. Also recognized as the Hydra-Antlia extension \citep{kraan1995,Kraan2000_vela}, the zoomed plot in the right panel reveals two branching structures crossing the GP at $\ell \sim 278^{\circ}$ and $282^{\circ}$ at $2000-3500$ \kms{} and $\, \ell \sim 280^{\circ}$ at $3500-5000$ \kms{}, concluding with a group of galaxies at ($\ell,~b) \sim (281^{\circ}, -7^{\circ}$). This feature is also evident as a prominent peak at $2000-3000$ \kms{} in the sensitivity curve plot in Fig. \ref{fig:sensitivity_curve_v2}.

In the higher velocity range of $5000-6500$ \kms{}, The alignment of blue dots in the wedge suggests the presence of a significant structure above the GP connecting to the Vela cluster AS639 ($280^{\circ}, +6^{\circ},{\sim}6000$ \kms{}; \citealt{Stein1996}), possibly originating from $ \ell \sim 265^{\circ}, cz \sim 5500$ \kms{}.\\

\noindent
$\mathit{6500} < cz \leq \mathit{11000} ~km\,s^{-1}$: The inclusion of new Vela$-$\hi{} data reveals previously undiscovered wall-like interconnecting structures (cyan dots) that cross the ZOA. The first hint of this structure was observed as a probable filament within the GP in \cite{Rajohnson2024} at $\ell \sim 260^{\circ}-277^{\circ}$ and velocity ranges around $7000-9000$ \kms{}.
In the on-sky plots, this is observed as a diagonal structure of cyan dots ($6500-8000$ \kms{}) crossing the inner GP within the longitudinal range of $266^{\circ} \leq \ell \leq 275^{\circ}$. It is reflected as the highest peak in the velocity histogram in Fig. \ref{fig:sensitivity_curve_v2}. This structure not only extends spatially below the GP but also appears to be connected to AS639 and higher velocity structures below the Plane (see orange dots in the velocity wedges). It then reaches the blue dots within $9500-11000$ \kms{} at $276^{\circ} \leq \ell \leq 280^{\circ}, b \sim -6^{\circ}$, which also appear to originate from the alignment of blue dots that cross the GP at $\ell \sim 281^{\circ}$. This crossing results in a distinctive void of approximately 10 Mpc in size at $\ell \sim 279^{\circ}, b \sim 0^{\circ}$ (see magenta circles in Fig. \ref{fig:LSS2_and_3D} and Fig. \ref{fig:LSS_wedge_ann}, labeled V1). \\

\noindent
$\mathit{11000} \leq cz \leq \mathit{15500} ~\text{km}\,\text{s}^{-1}$: Despite the lower density of data points for Vela$-$\hi{} compared to Vela$-$SMGPS, the continuation of green dots ($12500-14000$ \kms{}) suggests potential branching of three structures of the narrow filaments discovered in \cite{Rajohnson2024}. These structures cross the GP near longitudes of $\ell \sim 270^{\circ},276^{\circ}$, and $279^{\circ}$. However, due to the limited number of Vela$-$\hi{} detections, it is uncertain whether the first two potential crossings of the filaments extend above the GP (see thinner dashed purple lines). 
 
\subsubsection{Higher-$z$ Structures: $cz > 15500$ 
\kms{}}\label{sec5:higher_z_LSS}

\begin{figure*}
\centering
    \subfigure{\includegraphics[width=\textwidth]{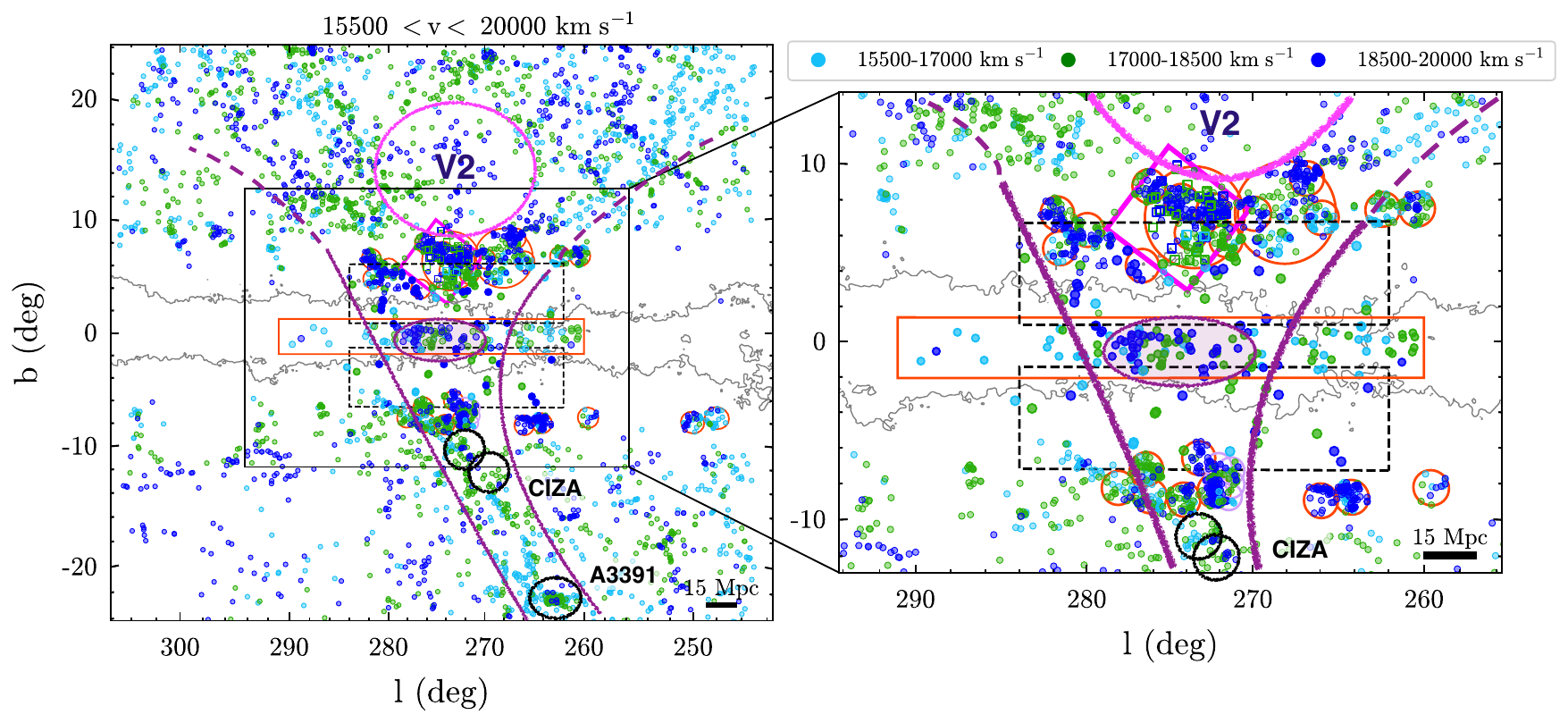}}\vspace{-1.5em}
    \hfill
    \subfigure{\includegraphics[width=\textwidth]{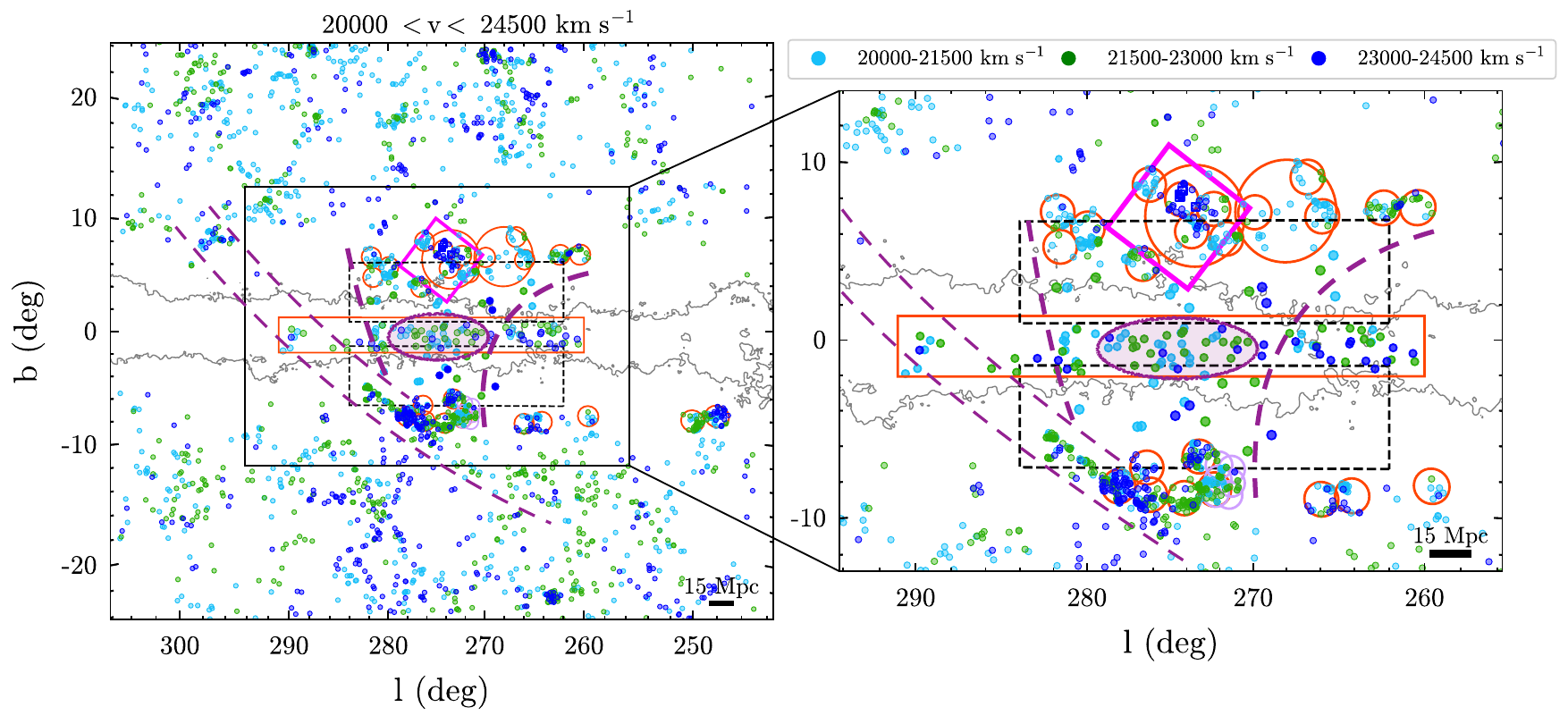}}\vspace{-1.5em}
	\caption{Similar to Fig. \ref{fig:LSS2_and_3D}, but featuring two panels for the high-redshift velocity range, specifically $15500 < cz < 24500$ \kms{}.}
	\label{fig:LSS2_and_3D_2}
\end{figure*}

Figure \ref{fig:LSS2_and_3D_2} presents the higher redshift ranges, which are directly relevant to our investigation of VSCL. The panels are divided into two segments ($15500-20000$ \kms{} and $20000-24500$ \kms{}) to provide insights into both VSCL walls. Due to the lower sensitivity, the links are not as clearly depicted compared to the lower redshift ranges. 

\noindent
$\mathit{15500} \leq cz \leq \mathit{20000} ~km\,s^{-1}$: galaxies marked by the continuation of green points ($17000-18500$ \kms{}) implies the presence of Wall 1 (W1). This structure appears to originate from a dense clump at ($\ell, ~b) \sim (272.5^{\circ}, 6^{\circ}$), intersecting the ZOA at least at around $\ell \sim 275^{\circ} \pm 1^{\circ}$, and reaching another concentration below the GP at ($\ell, ~b) \sim (276.5^{\circ}, -7^{\circ}$). The alignment of blue dots at $18500-20000$ \kms{}, on the other hand, seems to cross at least two locations in the GP (see purple-shaded circle). One branch likely merges in the inner GP at $\ell \sim 277^{\circ}$, while another filament of blue dots, originating from above the GP, crosses at $\ell \sim 272.5^{\circ}$, reaching the clustering below the Plane at ($\ell, ~b) \sim (272.5^{\circ}, -8^{\circ}$). The velocity distribution of Vela$-$\hi{} shown in Fig. \ref{fig:sensitivity_curve_v2} further supports these findings, displaying two peaks indicating overdensities at $17000-18000$ \kms{} and $19000-20000$ \kms{}, although with lower significance than in Vela$-$SMGPS distribution (see Fig. \ref{fig:hist_vel}).

\begin{landscape}
\begin{figure*}
    \centering
    \subfigure{\includegraphics[width=0.69\textwidth]{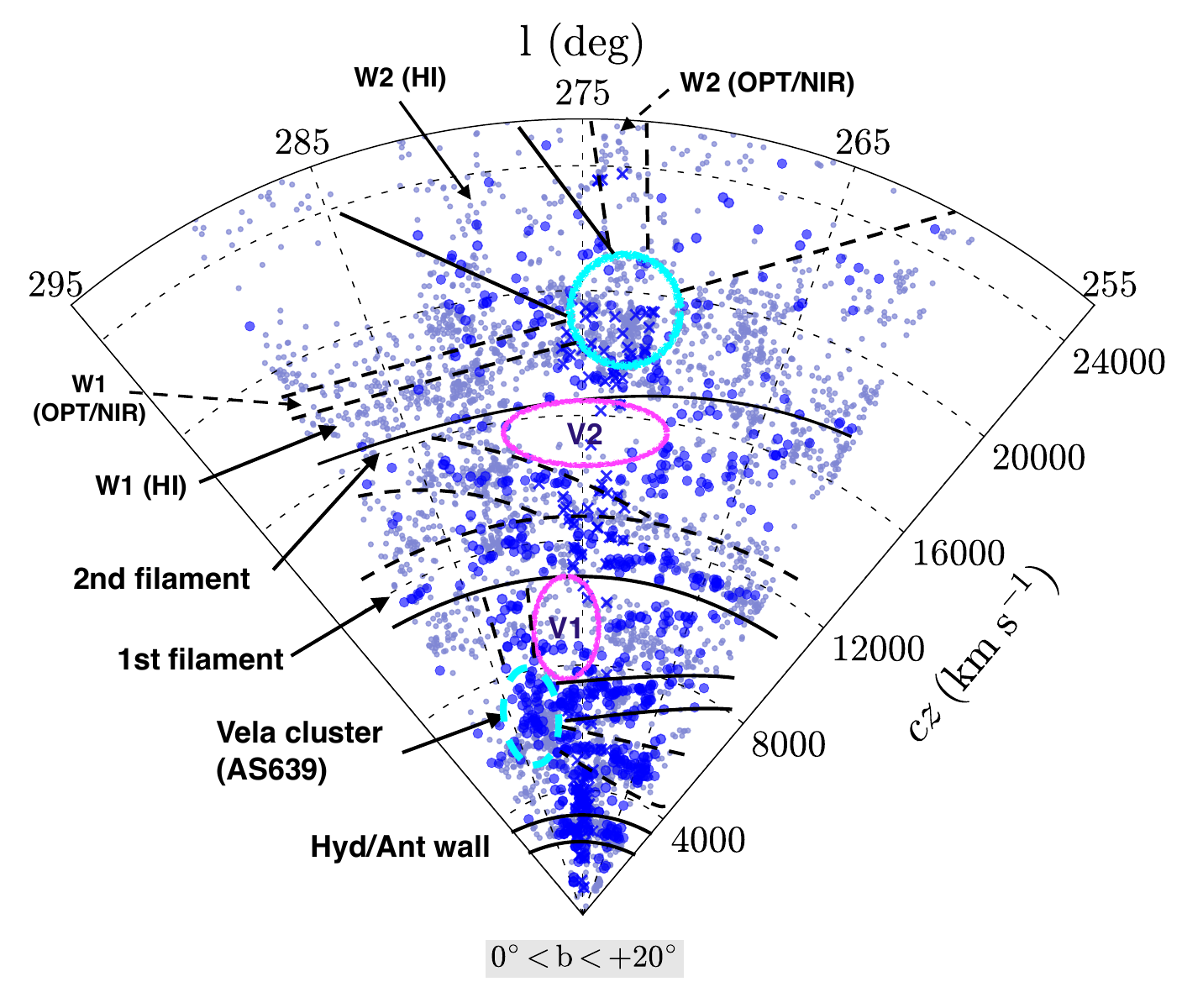}}
	\subfigure{\includegraphics[width=0.69\textwidth]{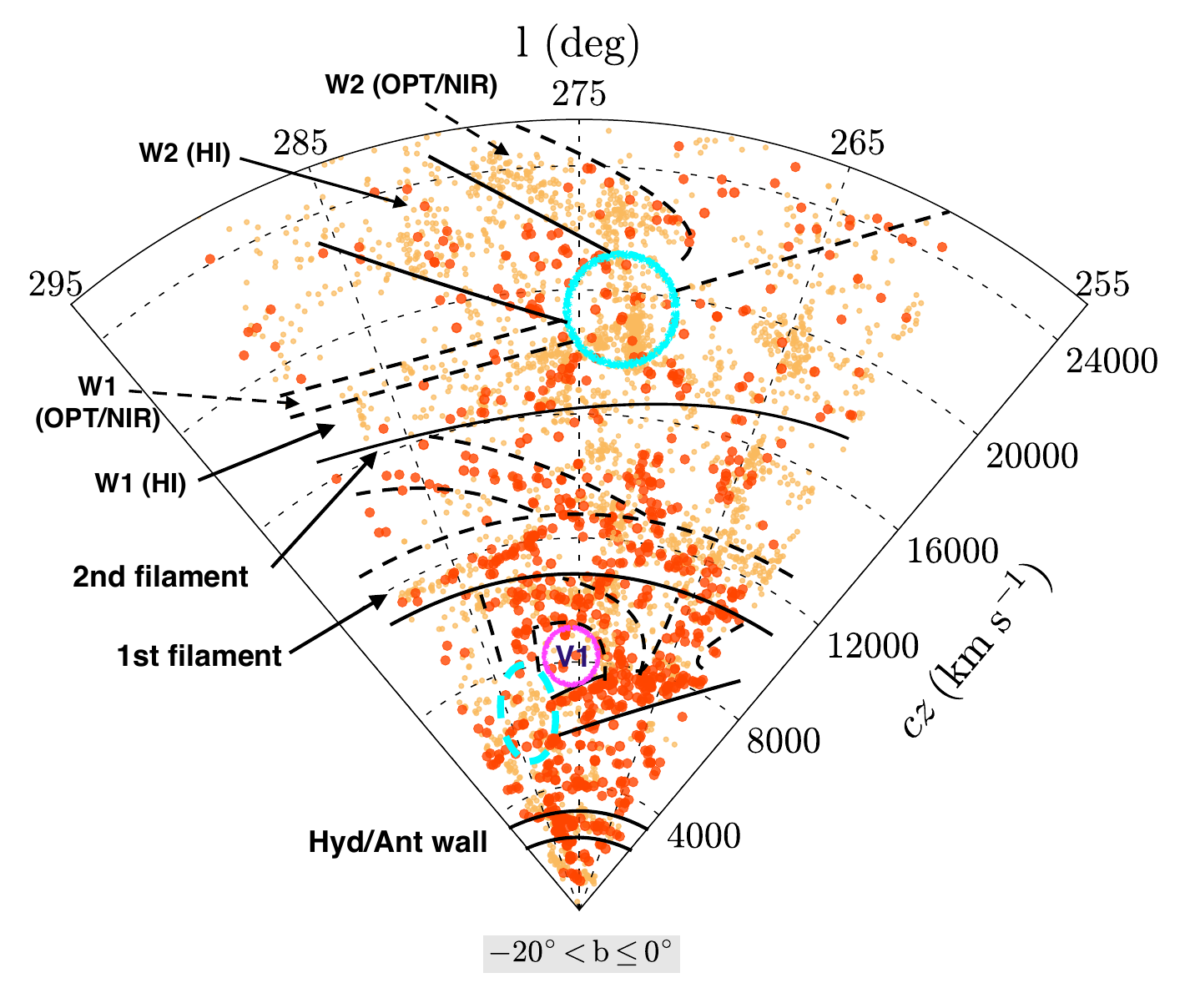}}\vspace{-1em}

    \vspace{1cm}
	\caption{Annotated redshift wedge diagrams illustrating galaxies located above the Plane in blue (left panel) and below the Plane in orange (right panel), covering the Galactic latitude range of $-20^{\circ} < b < +20^{\circ}$ and Galactic longitude range $260^{\circ} \leq \ell \leq 290^{\circ}$, with $cz < 25000$ \kms{}. The smaller, lighter points represent ancillary data from OPT and NIR surveys surrounding the surveyed area, while the larger dots exclusively denote \hi{} detections from Vela$-$SMGPS and Vela$-$\hi{}. Recent \hi{} detections from WALLABY-Vela are marked by blue crossed symbols in the left panel.}
	\label{fig:LSS_wedge_ann}
\end{figure*}
\end{landscape}

A notable observation in this panel is the connection from the cluster A3391 ($262.37^{\circ}, -25.16^{\circ}$, 15919 \kms{}), where cyan and green dots ($15500-18500$ \kms{}) extend to the clustering of blue dots ($18500-20000$ \kms{}) below the GP. This connection aligns with two X-ray clusters (CIZA J0812.5-5714 and CIZA J0820.9-5704; \citealt{Ebeling2002}), at approximately the same redshift (18587 and 18287 \kms{}, respectively). The structure then continues as a dense filamentary structure containing clusters and groups above the GP. This might explain why the two peaks at $17000-18000$ \kms{} and $19000-20000$ \kms{} are prominently observed above the Plane for Vela-\hi{} (cf. Fig. \ref{fig:hist_vel}). 
We propose that the underdensity visible above the GP at $\ell \sim 273^\circ$, $b \sim 15^\circ$, around 15500 \kms{}, just before the redshift of the first wall, is a genuine void (designated as V2) due to the low number of galaxies shown by the sensitivity curve in Fig. \ref{fig:sensitivity_curve_v2}. Moreover, the wedge diagram in the left panel of Fig. \ref{fig:LSS_wedge_ann} shows hardly any galaxies in the region above the GP, out to $0^{\circ} < b < 20^{\circ}$. 
The overall arrangement of the two walls in this velocity range appears to be part of a multi-branching structure that coincidentally intersects within the longitude range of $270^{\circ} \leq \ell \leq 279^{\circ}$ and around $b \sim 0^{\circ}$. \\

\noindent
$\mathit{20000} \leq cz \leq \mathit{24500} ~km\,s^{-1}$: Despite the low number of detections due to sensitivity drop, a hint of a broad overdensity around $20000-21500$ \kms{} (cyan dots) and $21500-23000$ \kms{} (green dots) is observed within the range $270^{\circ} \leq \ell \leq 279^{\circ}$. It is slightly broader than suggested from the suspected Vela$-$SMGPS data. The structure seems to align with the X-ray cluster CIZA J0745.1-5404/ESO 163-IG015 ($266.84^{\circ}, -14.36^{\circ}, 22185$ \kms{}; \citealt{Kocevski2007}). It reinforces the findings from Vela$-$SMGPS that the crossing seems to occur in the inner GP around velocities of $18500-21500$ \kms{}. This potential intersection corresponds to the rough estimate that places the center of mass overdensity of the VSCL at the lowest latitudes around approximately ${\sim}272^{\circ}$ (also see the cyan circle in the wedge diagram). This location was suggested from the examination of the region near Vela$-$OPT/NIR in \citetalias{Kraan-Korteweg2017}. The result also aligns with the general direction of the core of a significant overdensity, as suggested from independent density and velocity field reconstructions by \cite{Sorce2017} and \cite{Courtois2019}. This was recently strengthened by results from the linear density field reconstruction of the WALLABY-Vela \citep{Mould2024}. They place the core of the VSCL close to the center of mass overdensity predicted by \citetalias{Kraan-Korteweg2017} with a $4\sigma$ significance. 

The wedge diagrams in Fig. \ref{fig:LSS_wedge_ann} furthermore reveal distinguishable features between the walls traced by \hi{} detections and those traced by OPT/NIR detections, suggestive of possible offsets between them. The shift of the \hi{} detections with respect to OPT/NIR walls could indicate an ongoing collapse of the \hi{}-rich galaxies towards the walls of the VSCL. Galaxies from the surrounding cosmic web could be gravitationally attracted and falling towards these walls, making those located at the periphery more \hi{}-rich. Alternatively, interactions and pre-processing in the filaments may have stripped gas from the OPT/NIR wall \citep[e.g.,][]{Sarron2019}. \\ 

The less pronounced supercluster signature in Vela$-$\hi{} compared to Vela$-$SMGPS may not only be due to its lower sensitivity. It is plausible that the relatively high number of detections concentrated in Vela$-$SMGPS is caused by the signature of the walls crossing at these lowest latitudes, raising -- nearly doubling -- the galaxy counts. While Vela$-$\hi{}, with its 0.75 \mJy{} sensitivity, is `only' revealing the most \hi{}-massive galaxies in Vela, they do trace the walls, but because of their lower density, the walls may overall contain fewer very rich \hi{} galaxies as they stream towards VSCL. This interpretation aligns with \citetalias{Kraan-Korteweg2017}, proposing that the VSCL is a supercluster in formation, characterized by two merging wall-like structures. This hypothesis would also bolster the expectation suggested in \cite{Einasto2007} that high-density clusters are connected by lower-density filaments. In combination with the lower density, this could explain the low number of detections along the walls of VSCL. 
The spatial arrangement of galaxies evident in both the on-sky plot and wedge diagrams strongly suggests that the VSCL possesses the characteristic morphology of a rich supercluster. This morphology is characterized by an elongated, wall-like structure and the presence of multi-branching filaments, as proposed by \citep{Einasto2011b}. To get further insight into the crossing, a deeper \hi{} survey targeting the suggested core location of the VSCL is deemed essential for further validation. Alternatively, running the DisPerSE filament finder \citep{Sousbie2011} might also help identify persistent structures quanti-

\onecolumn
\twocolumn

\noindent
tatively.

\section{Conclusions}\label{sec5:conclusion}

In our quest to unravel the morphology and characteristics of the Vela Supercluster, concealed within the ZOA, we conducted a blind \hi\ systematic MeerKAT Vela Supercluster survey, denoted as Vela$-$\hi{}. This survey was strategically positioned to address the formerly unexplored patches at $1^{\circ} \leq b \leq 6.2^{\circ}$ and $-6.7^{\circ} \leq b \leq -2^{\circ}$ in the Vela region. The survey design aims to establish connections between Vela$-$SMGPS ($-2^{\circ} \leq b \leq 1^{\circ}$, \citealt{Rajohnson2024}), and deep OPT/NIR observations at the outer edges of the ZOA in the general VSCL direction ($5^{\circ} \leq |b| \leq 10^{\circ}$, \citetalias{Kraan-Korteweg2017}). Over approximately 67 hours, 667 MeerKAT pointings were collected, each lasting 5 minutes, covering an extensive area (${\sim}242$ deg$^2$) both above and below the GP at $263^{\circ} \leq \ell \leq 284^{\circ}$. 
The dataset was analyzed and reduced, encompassing a velocity range up to $cz \sim 25000$ \kms{}.\\

The Nyquist sampling configuration reduced the initial rms during data acquisition (0.88$-$9.27 \mJy{}) to a final noise level of 0.68$-$1.24 \mJy{} after solar RFI flagging and mosaicking were conducted.

In total, 719 \hi\ sources in Vela$-$\hi{} were detected and parameterized, revealing 508 previously unknown galaxies. Only 211 of these detections were previously recorded in the literature. Among them, only 66 have available redshift information with just five over $cz > 16000$ \kms{}. These figures underscore the efficacy of blind \hi{} surveys in identifying obscured galaxies within the ZOA. Obtaining the redshifts of these galaxies would be arduous or nearly impossible in optical, particularly at higher redshifts (VSCL distance). The Vela$-$\hi{} parameters and positions demonstrate high quality: the measured integrated fluxes, linewidths, and systemic velocities are highly consistent with the previous Vela$-$SMGPS and HIZOA surveys. \\

The majority of OPT/NIR cross-matches lie within $\ang{;;5.6} \pm \ang{;;3.4}$ of the \hi{} positions. Considering that Vela$-$\hi{}’s internal coordinate precision is approximately $\ang{;;6}$, this demonstrates excellent positional accuracy. The observed difference of $70 \pm 51$ \kms{} between the optical and \hi{} velocities is also within the uncertainty of the optical velocities, which is on the order of 100 \kms{}.

The inclusion of 719 new redshift measurements significantly solidifies our understanding of the overall picture of the LSS when linking previously known structures at much higher latitudes in the Vela region. It clearly confirms the continuation of the Hydra/Antlia extension across the GP. Additionally, we have uncovered previously unknown LSS, including a diagonally structured wall possibly linked to AS639 and a void with a diameter of approximately 10 Mpc. At the distance of the VSCL, the distribution, connected to known clusters, reinforces the multi-branching morphology of the supercluster and affirms the existence of two potentially merging walls. This is further supported by recent independent reconstructions. When looking at the wedge diagrams, apparent offsets between the \hi{} and OPT/NIR walls are observed, which seems to support the notion that VSCL is a supercluster in formation. \\

To conclude, combining data from Vela$-$\hi{} and Vela$-$SMGPS has provided a clearer picture of the extent of the VSCL and improved our understanding of LSS in the Vela region out to 25000 \kms{}. The presence of this previously hidden core, supported by \hi{} observations prominent both above and below the GP, is a first significant step toward assessing its effect on the cosmic flow fields. However, quantifying the core and its impact requires more extensive studies beyond the scope of a single paper. We are addressing this in an upcoming paper, where we aim to measure the potential overdensity at the core of the VSCL using \hi{} mass functions, and we have signs of a clear quantitative overdensity. With adequate photometry, the data quality and statistics should also enable us to determine peculiar velocities and potentially obtain velocity field distributions in the Vela region in the future.

\section*{Acknowledgements}
The MeerKAT telescope is operated by the South African Radio Astronomy Observatory, which is a facility of the National Research Foundation, an agency of the Department of Science and Innovation. We acknowledge the use of the ilifu cloud computing facility -- \url{www.ilifu.ac.za}, a partnership between the University of Cape Town, the University of the Western Cape, the University of Stellenbosch, Sol Plaatje University, the Cape Peninsula University of Technology and the South African Radio Astronomy Observatory. The ilifu facility is supported by contributions from the Inter-University Institute for Data Intensive Astronomy (IDIA – a partnership between the University of Cape Town, the University of Pretoria, the University of the Western Cape and the South African Radio astronomy Observatory), the Computational Biology division at UCT and the Data Intensive Research Initiative of South Africa (DIRISA). This work made use of the CARTA (Cube Analysis and Rendering Tool for Astronomy) software (DOI 10.5281/zenodo.3377984 – \url{https://cartavis.github.io}).

This paper makes use of the MeerKAT data with Project ID: SCI-20210212-SR-01. (Part of) the data published here have been reduced using the CARACal pipeline, partially supported by ERC Starting grant number 679629 ``FORNAX'', MAECI Grant Number ZA18GR02, DST-NRF Grant Number 113121 as part of the ISARP Joint Research Scheme, and BMBF project 05A17PC2 for D-MeerKAT, and partially supported by the South African Research Chairs Initiative of the Department of Science and Technology and National Research Foundation. Information about CARACal can be obtained online under the URL: \url{https://caracal.readthedocs.io/en/latest/}.

SHAR, RCKK, HC, NS, SK, and DJP are supported by the South African Research Chairs Initiative of the Department of Science and Technology and National Research Foundation. 

We acknowledge the usage of the HyperLeda database (\url{http://leda.univ-lyon1.fr}). This research has made use of: the NASA/IPAC Extragalactic Database (NED), which is operated by the Jet Propulsion Laboratory, California Institute of Technology, under contract with the National Aeronautics and Space Administration; the NASA’s Astrophysics Data System Bibliographic Services; the VizieR catalogue access tool, CDS, Strasbourg, France (DOI:10.26093/cds/vizier). The original description of the VizieR service was published in 2000, A\&AS 143, 23. This publication makes use of data products from: the Two Micron All Sky Survey, which is a joint project of the University of Massachusetts and the Infrared Processing and Analysis Center/California Institute of Technology, funded by the National Aeronautics and Space Administration and the National Science Foundation; the Wide-field Infrared Survey Explorer, which is a joint project of the University of California, Los Angeles, and the Jet Propulsion Laboratory/California Institute of Technology, funded by the National Aeronautics and Space Administration. This research made use of \href{http://www.astropy.org}{Astropy}, a community-developed core Python package for Astronomy \citep{astropy:2013, astropy:2018}.

\section*{Data Availability}

The full Vela-\hi{} catalog and atlas can now be accessed online. The former is available as supplementary ASCII material, while the latter can be found in the Zenodo repository at \url{https://doi.org/10.5281/zenodo.12522807}. The raw MeerKAT Open Time data should now be publicly accessible at \url{https://archive.sarao.ac.za}, and the author can provide reduced \hi{} cubes and/or mosaics upon request.



\bibliographystyle{mnras}
\bibliography{reference} 



\newpage
\pagebreak
\onecolumn 
\appendix

\section{Residual RFI}\label{sec5:RFI}

This appendix explains the method employed to address short-track bugs and solar RFI observed in the Vela$-$\hi{} datasets before executing the CARACal pipeline. 

\subsection{Short track bugs}\label{sec5:bug}
The short track bug occurs during telescope slewing, where the initial two or three integrations from this motion are erroneously recorded as a track lasting eight seconds or less when the telescope moves to a new field. In Fig. \ref{fig5:bug}, we have highlighted these short tracks in yellow. They are visibly shorter (see the column \textsc{Timerange (UTC)}), containing fewer rows of data compared to an actual track. If left unflagged, the short track bug can cause emission from the previous field to erroneously appear in the subsequent field as faint `ghost' images. While \texttt{CARACal} automatically identifies and flags this bug, it can also be manually addressed in \texttt{CASA} \citep{CASA2022} 
by carefully noting the affected scans and flagging them before calibration. 

\begin{adjustbox}{center,caption={\small An example of a \textsc{casa} `listobs' file summarizing the observation information for the measurement set of block C2. Short-track bugs, which require flagging, have been highlighted in yellow.},
label={fig5:bug},nofloat=figure,vspace=\bigskipamount}
\includegraphics[width=\linewidth]{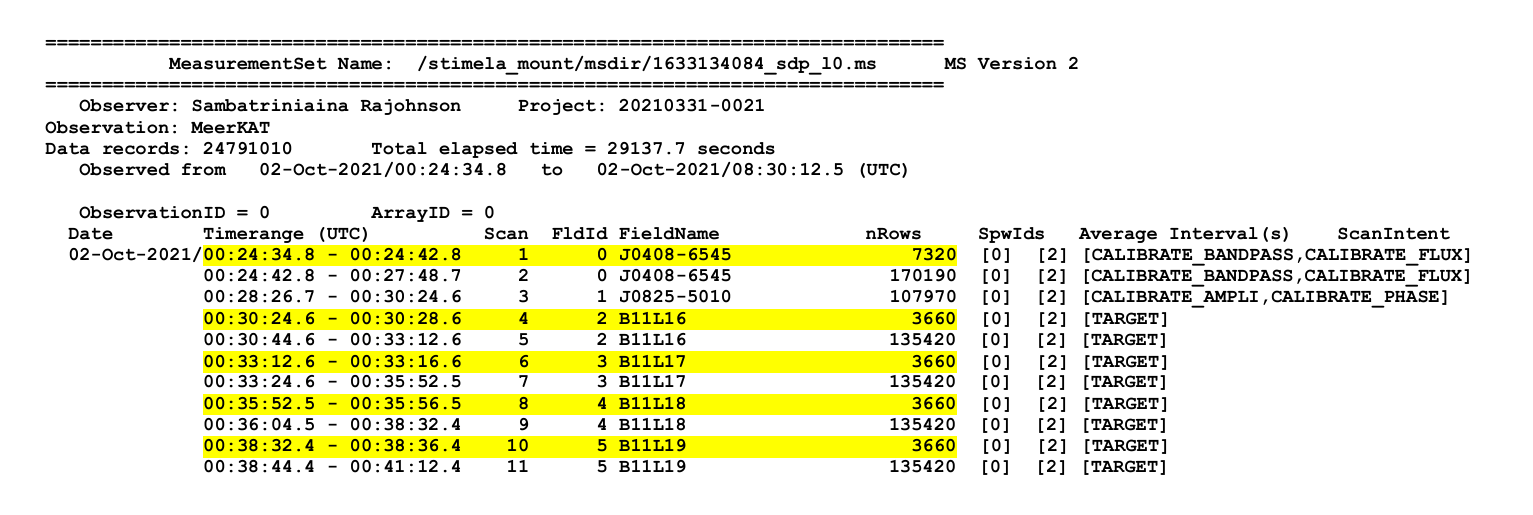}
\vspace*{-6mm}
\end{adjustbox}

\subsection{Solar RFI and 1380 MHz spike}\label{sec5:RFI_sol}
Initially, our blocks of observations all started during nighttime. However, due to the extensive duration ($8-9$ hrs) of each session and a large number of targets, roughly 69\% of our data was acquired during the daytime. This distribution is summarized in Table \ref{tab:flagged_data}, where blocks A1 and A2 have collected more than 80\% of their data during daytime. This had significant implications, particularly for short baselines directed toward the Sun during sunrise and sunset, which were susceptible to picking up solar interference. \\

\begin{figure*}
    \centering    
    \includegraphics[width=\linewidth]{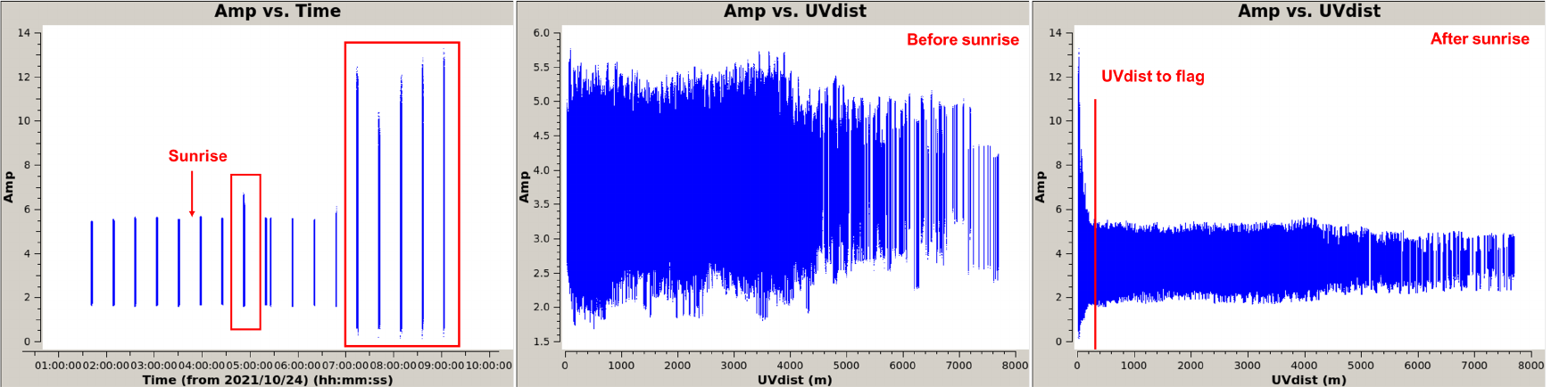}
    \caption{Example of diagnostic plots for the gain calibrator J0825-5010 prior cross-calibration. The left panel displays the amplitude variation over time during the observations, with the sunrise (03:46 UTC) indicated by a red arrow. Scans impacted by solar RFI are marked by red boxes. The middle and right panels showcase the amplitude variation with UV distance, both pre and post-sunrise, respectively. In the right panel, the maximum UV distance for flagging (UVdist < 1.5 k$\lambda$ or 315m) after sunrise, specifically for the affected baselines, is denoted by the red vertical line.}
    \label{fig:inspection}
\end{figure*}

\begin{figure}
	\centering
	\subfigure{\includegraphics[width=0.5\linewidth]{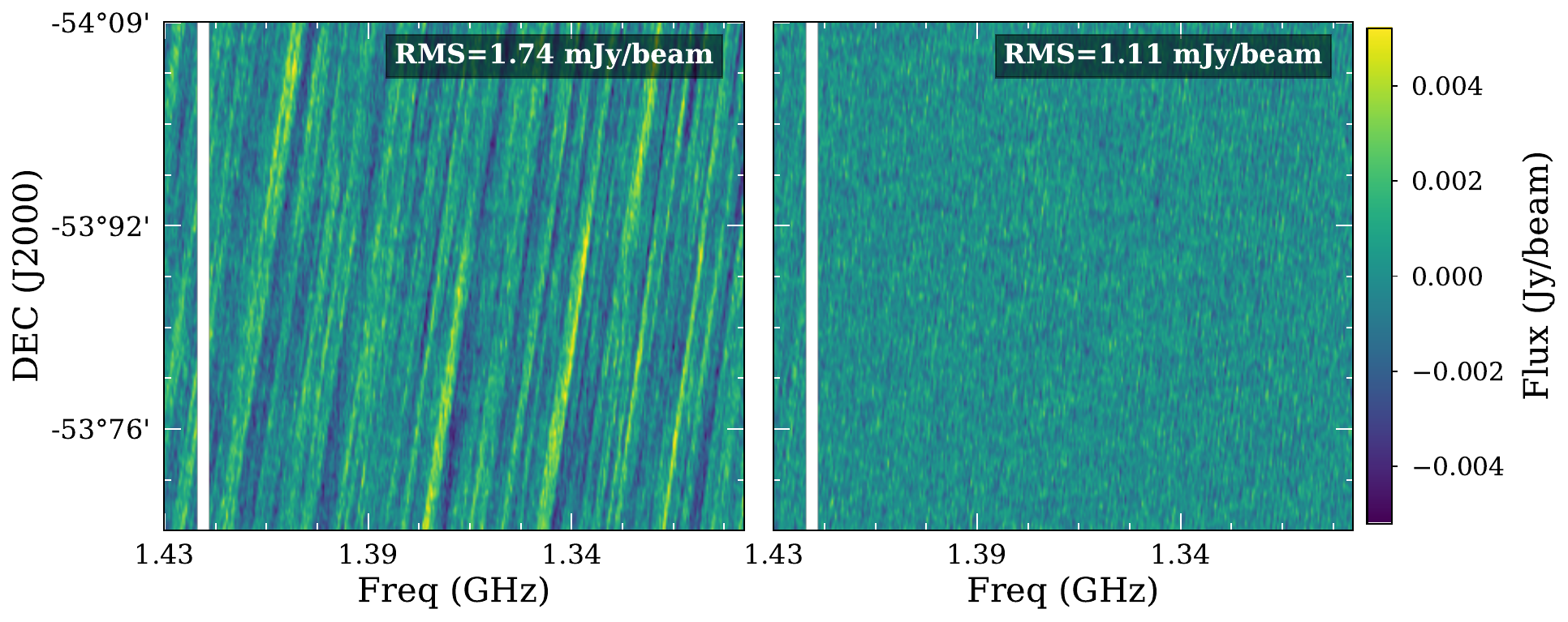}}\vspace{-1.5em}
	\subfigure{\includegraphics[width=0.5\linewidth]{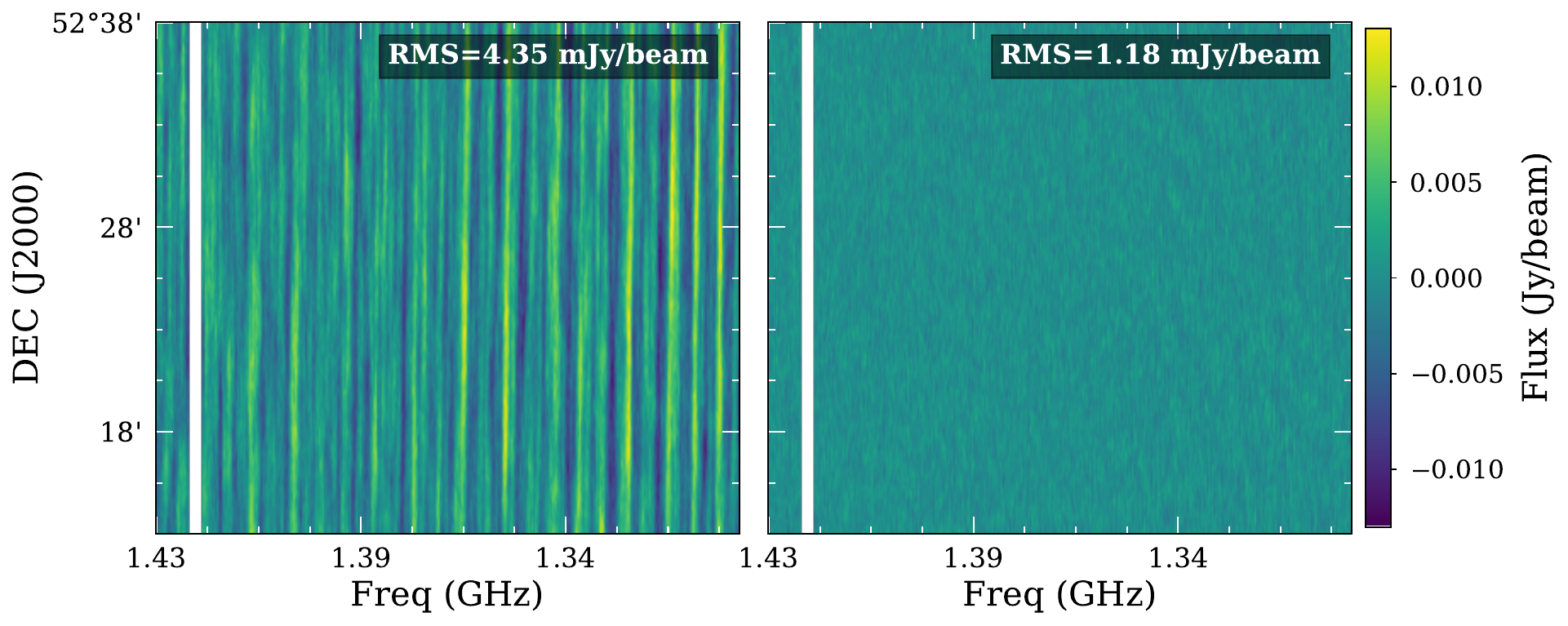}}
	\centering
	\caption{Example of PV-slice of \hi{} cubes from two target fields. The left panel demonstrates the presence of ripple artifacts, which are effectively removed in the right panel following the solar RFI flagging process.}
	\label{fig:solar_rfi}
\end{figure}

Upon inspecting the data, we observe a sudden fluctuation in amplitude (see left panel of Fig. \ref{fig:inspection}). The observed deviation, occurring immediately after sunrise, is attributed to solar RFI. 
The latter translates into peculiar ripples discernible in both the spatial and spectral axes when generating the \hi\ cubes, as demonstrated in the left panels of Fig. \ref{fig:solar_rfi}.

As shown in the middle panel of Fig. \ref{fig:inspection}, the variation of amplitude as a function of UV distance remains stable before sunrise. However, after sunrise, a significant variation is observed, particularly for UV distances below 300 m. When this occurs, it is crucial to flag affected daytime visibilities. It is important to emphasize that not all daytime data requires flagging. Specifically, flagging is necessary during the interference duration and for the affected baselines only. The baselines primarily affected are those with the shortest projected lengths when aligned with the rising or setting Sun’s path. The most severely affected baselines are listed in the MeerKAT Users portal documentation\footnote{\url{https://skaafrica.atlassian.net/wiki/spaces/ESDKB/pages/336232568/Known+issues} (status 28/08/2024, 14:05 UTC)}. Our measurements, detailed in Table \ref{tab:flagged_data}, indicate that baseline lengths shorter than 315 m (${\sim}1.5$ k$\lambda$) should be flagged for blocks A2, B2, C2, D1, and D2. For A1 and B1, this threshold is 210 m (${\sim}1$ k$\lambda$). The flagged data and the total number of affected baselines together comprise less than 8.1\% of the total baselines and data, resulting in a relatively low impact on sensitivity.

\clearpage
\begin{table}
    \small
	\centering
        \setlength{\tabcolsep}{4pt}
	\caption{Overview of flagged Solar RFI in Vela$-$\hi{} data, including the percentage of daytime data and the affected baseline length that requires flagging}
	\label{tab:flagged_data}
		\begin{tabular}{c c c c }
		    \hline
			\hline
			 Block ID & Daytime data& Affected baseline length& Flagged data\\
                  & (\%)& (m)& (\%)\\
			\hline
         A1& 81.4& 210& 7.9\\
         A2& 89.2& 315& 12.0\\
         B1& 64.9& 210& 5.7\\
         B2& 69.1& 315& 6.1\\
         C1& 64.5& 210& 6.2\\
         C2& 40.9& 315& 6.1\\
         D1& 75.2& 315& 10.7\\
         D2& 71.9& 315& 10.4\\
            \hline
		\end{tabular}
	\end{table}
	
\begin{figure}
	\centering
	\subfigure{\includegraphics[width=0.625\linewidth]{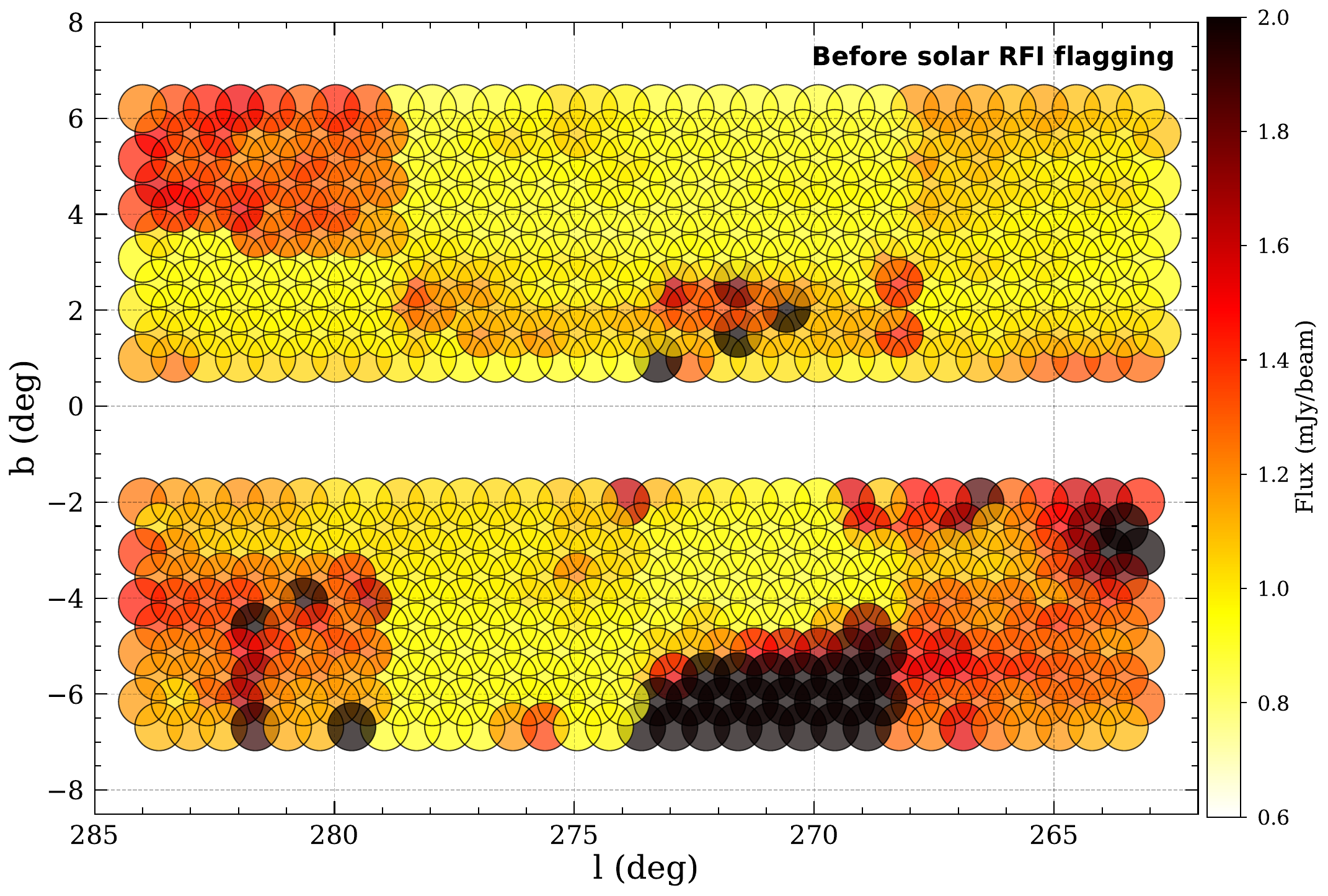}}\vspace{-1.5em}
	\subfigure{\includegraphics[width=0.625\linewidth]{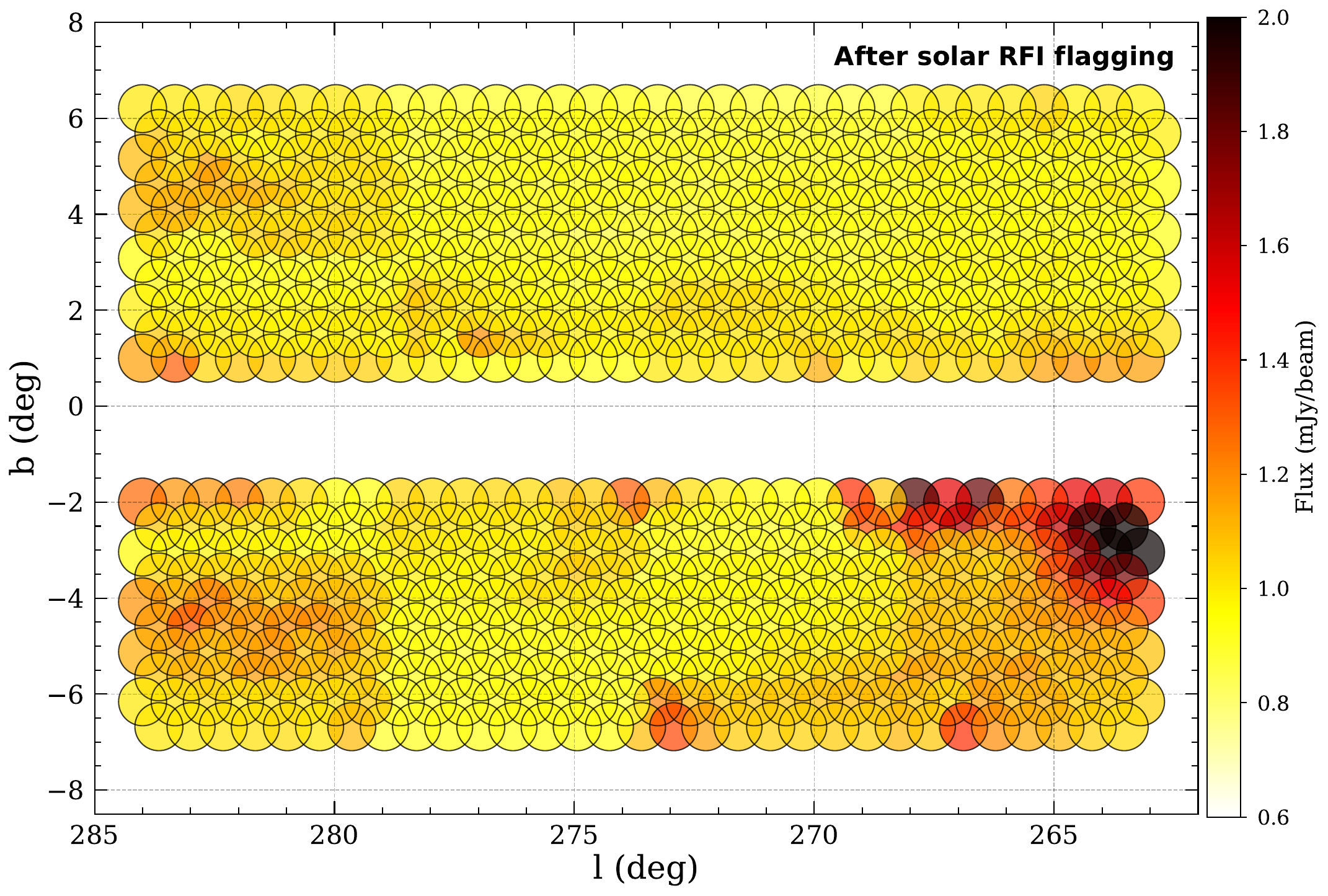}}
        \subfigure{\includegraphics[width=0.625\linewidth]{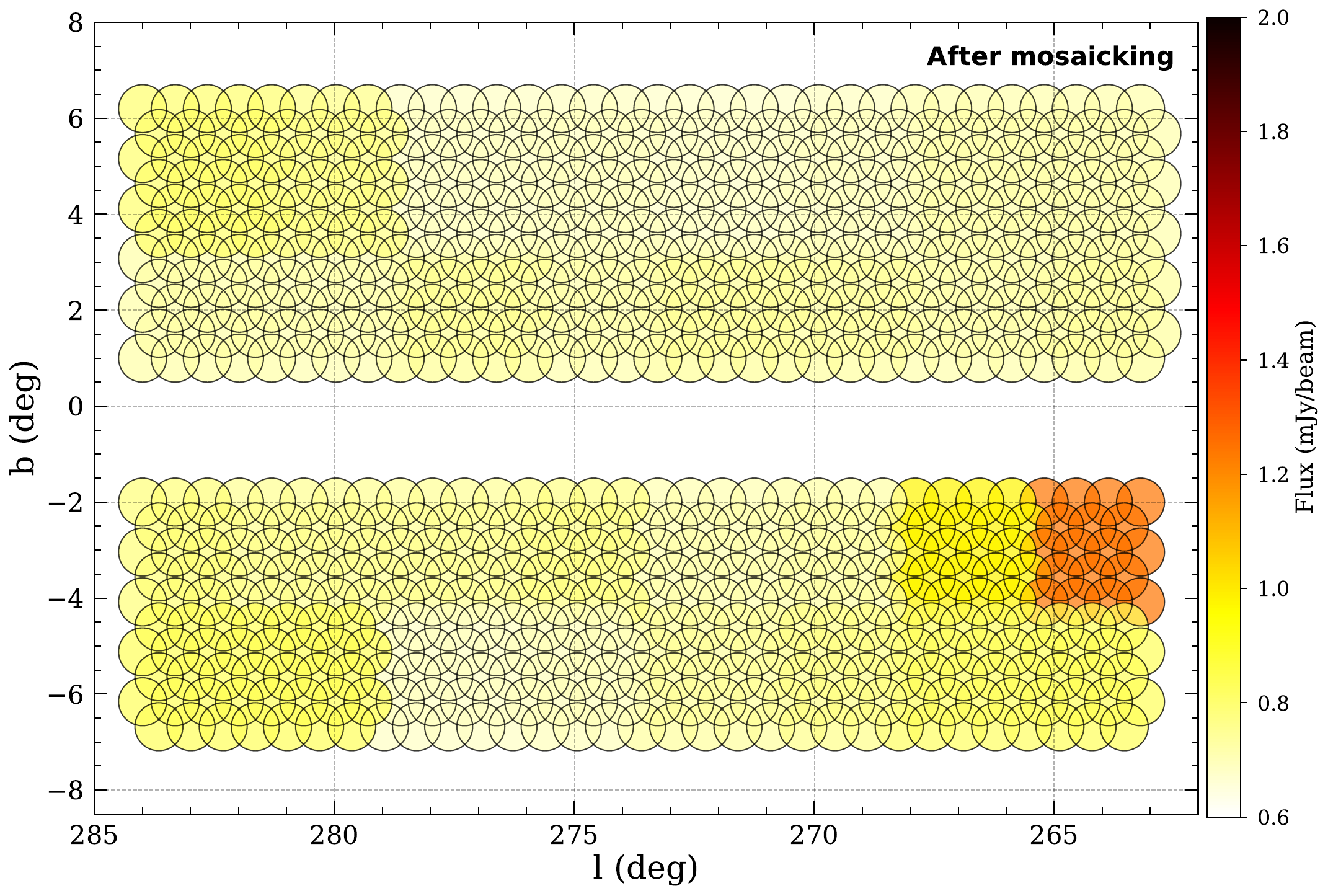}}
	\centering
	\caption{Rms variations and improvements per field before and after RFI flagging and mosaicking. In the bottom panel, all fields in the same mosaic share the same color code, indicating the mean noise of that mosaic.}
	\label{fig:rms_variations}
\end{figure}

After flagging (right panel of Fig. \ref{fig:solar_rfi}), a substantial reduction or complete elimination of the ripple effect in the position-velocity (PV) slice of the \hi\ cubes is observed, leading to a significant improvement in sensitivity. This noise improvement in the \hi{} cubes is further illustrated in the first two panels of Fig. \ref{fig:rms_variations}. The measured rms prior to solar RFI flagging ranges from $0.88$ to $9.27$ \mJy{}. Cubes previously highlighted in dark red (with rms $> 2$ \mJy{}) in the top panel, are now shaded in orange and yellow (i.e., rms ${\sim}1$ \mJy{}), marking a twofold increase in sensitivity after solar RFI flagging, with rms values now ranging from $0.88-2.57$ \mJy{}. This noise improvement continues to be remarkable after mosaicking (see Section \ref{sec5:mosaicking}), achieving an rms noise spanning from $0.68$ to $1.24$ \mJy{} in the bottom panel of Fig. \ref{fig:rms_variations}. However, we note that the color coding in the bottom panel does not account for the spatial variation of the noise within each mosaic. It only represents the mean noise in the mosaic, so all fields belonging to the same mosaic (as shown in Fig. \ref{fig:blocks_mosaic}) are assigned the same color and, hence, the same noise level in this figure.\\

Finally, an additional flagging step is performed after self-calibration and continuum subtraction, just before the imaging step. This step utilizes \textsc{casa} tasks \textsc{tfcrop} and \textsc{rflag} to eliminate residual RFI sidelobes, primarily originating from Global Positioning System (GPS) satellites. While the telescope system automatically flags the strong peak at ${\sim}$1380 MHz, further action is required if it persists in the spectral axis of the \hi\ cube. This scenario arises for blocks A2, B2, C1, and C2, so we applied time and frequency cutoffs of $4\sigma$ and 3$\sigma$ in \textsc{tfcrop}, and 4$\sigma$ in both time and frequency for \textsc{rflag}.

\begin{figure}
    \centering
    {\includegraphics[width=0.9\linewidth]{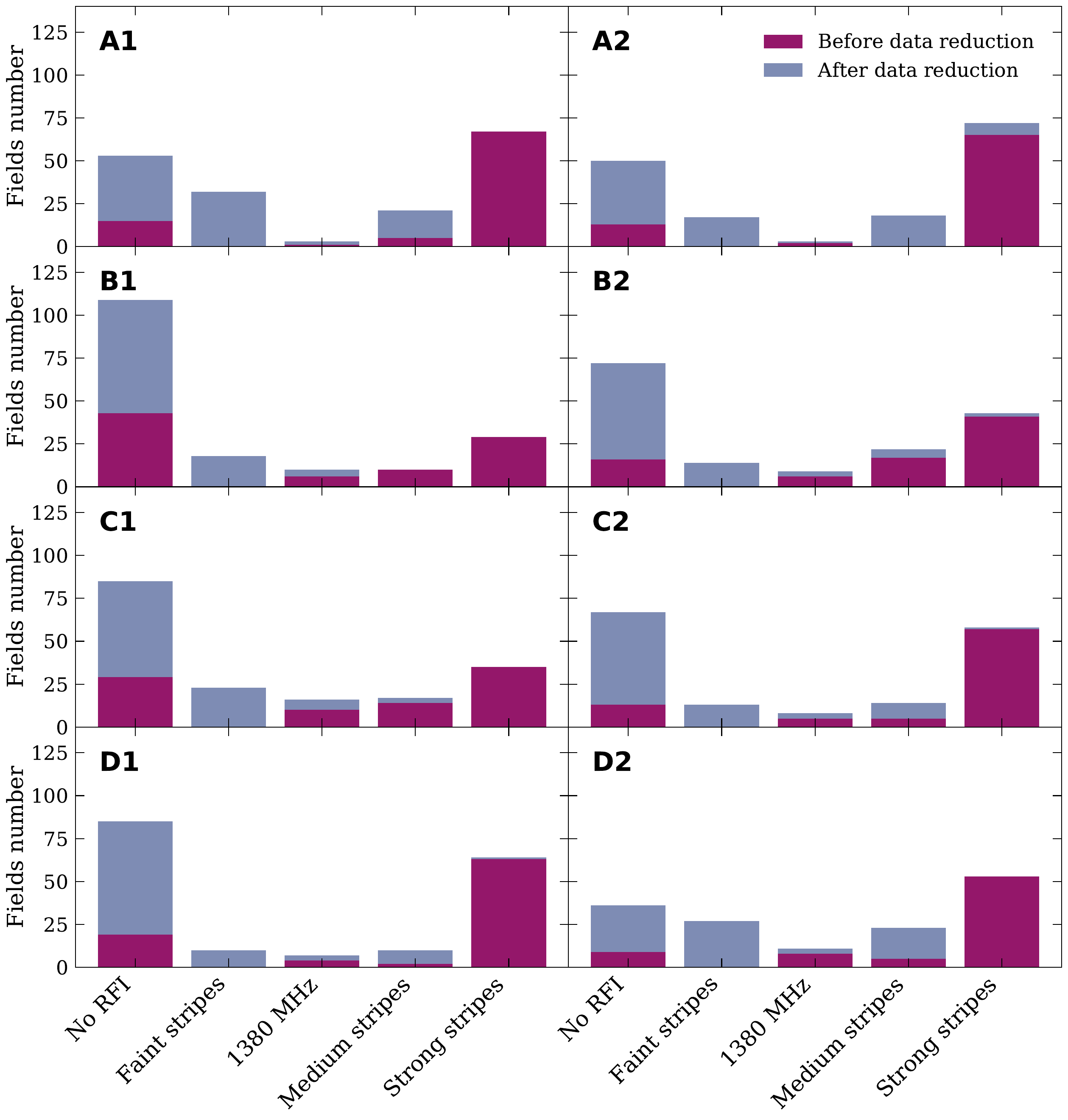}}
    {\caption{Histogram displaying the transition of fields from showing strong stripes affected by solar and 1380 MHz RFI to a state of no RFI in each Vela$-$\hi{} block, both before (light purple) and after (dark purple) data reduction \label{fig:improvement_all_block}}} \vspace{-1.5em}
\end{figure}

Prior to calibration and flagging, over half of the target fields in each block (greater than 50 fields) displayed some form of stripes (classified as either strong, medium, faint, or 1380 MHz spike) in their maps, impacting our observations. After solar RFI removal and data reduction, a significant improvement in data quality is observed. Now, the majority of fields are free from RFI or exhibit medium and faint stripes, as indicated by the light purple shading in Fig. \ref{fig:improvement_all_block}. While a few instances persist, they are considered acceptable.

\pagebreak
\onecolumn 

\section{Vela$-$\hi{} full catalog}\label{sec5:full_cat}

The following Appendix provides a detailed description of the catalog parameters for Vela$-$\hi{}, which are primarily sourced from the SoFiA output unless mentioned otherwise. The complete catalog is available in Table \ref{tab:cat}.

\begin{itemize}
    \item[] Column (1): Source Identifier (VSCL$-$\hi{} Jhhmmss-ddmmss) representing rounded right ascension and declination in equatorial coordinates (J2000) of the centroid position.
    \item[] Column (2): Indicates the name of the mosaic to which the source belongs. It is presented in the format of the block name along with the mosaic configuration (AXB), e.g., B1C [A1A to D2D].
    \item[] Columns (3) and (4): Galactic Coordinates ($\ell$ and $b$) of the centroid position in degrees.
    \item[] Column (5): Peak Flux density measured from the Global \hi{} profile in Jy. 
    \item[] Columns (6) and (7): Integrated Flux along with its corresponding error $\varepsilon_s$ in a unit of Jy \kms{}. This error value is the average standard deviation of the integrated fluxes, which are measured from four emission-free regions around the source. 
    \item[] Column (8): Mean Local rms in \mJy{}, averaged from four emission-free regions around the detection. 
    \item[] Column (9): Optical Heliocentric systemic velocity ($V_{\mathrm{hel}}$) in \kms{}. 
    \item[] Columns (10) and (11): Linewidths ($w_{20}, w_{50}$) in \kms{}, corrected for instrumental resolution, and representing the widths at which the profile achieves 20\% and 50\% of its peak flux density.
    \item[] Column (12): Logarithm of \hi{} Mass [$\log(M_{\mathrm{HI}}/\mathrm{M_{\odot}}$)]. No corrections for Local Group barycenter or instrumental resolution were applied.
    \item[] Column (13): Flags denoting the classification of detections. A flag of 1 indicates a solid detection, while 2 signifies a possible detection. Additional notations include an asterisk (*) alongside the flag, indicating that the detection was also identified in the overlapping mosaic. A double asterisk (**) denotes that the detection was found in more than two adjacent mosaics, and an exclamation mark (!) indicates a common detection with Vela$-$SMGPS.
    \item[] Column (14): Additional Notes and Multi-Wavelength Counterparts (if available). The notes cover various attributes: 
    \begin{itemize}
        \item[-] \emph{B$=\ang{;;40}$:} The detection's moment maps have been smoothed to a specific angular resolution (e.g., $\ang{;;40}$).
        \item[-] \emph{BL:} the Global \hi{} profile shows a wiggly baseline
        \item[-] \emph{e:} detection located near the mosaic edge.
        \item[-] \emph{i:} with signs of interactions or mergers.
        \item[-] \emph{r:} the detection is located in (or close to) a noisy area due to the ripple caused by residual solar RFI.
        \item[-] \emph{w.comp(s):} galaxy with one (or several) companion(s).
        \item[-] \emph{$w_{20}$ unc.:} $w_{20}$ measurement unclear due to baseline behaviour.
    \end{itemize}
    
    Additionally, galaxies detected in major \hi{}, OPT, and NIR catalogs are also indicated in the notes column. The methodology for identifying multi-wavelength counterparts is explained in Section \ref{sec5:multi_counterparts}. For galaxies found in HIZOA \citep{Staveley2016} and/or Vela$-$OPT/NIR (\citetalias{Kraan-Korteweg2017}, Kraan-Korteweg priv. comm), their object names (e.g., J0847-46, HyA0009, Vel0617) are provided. There are also other cases such as:
    \begin{itemize}
        \item[-] \emph{H:} counterpart from the southern region of the HIPASS catalog \citep{Meyer2004}.
        \item[-] \emph{I:} counterpart from the IRAS Point Source Catalog \citep{Helou1988}.
        \item[-] \emph{M:} counterpart from 2MASX \citep{Jarrett2000} and other 2MASS detections (\citealt{Skrutskie2006}, in the absence of 2MASX).
        \item[-] \emph{W:} counterpart from the Wide-field Infrared Survey Explorer catalog (WISE; \citealt{Wright2010,Cutri2014}).
        \item[-] \emph{Z:} counterpart from deep NIR follow-up observations conducted using IRSF \citep{Williams2014,Said2016}.\\
    \end{itemize}
 
\end{itemize}

\clearpage
\onecolumn 
\begingroup
{
\begin{landscape}
\fontsize{6.5pt}{9pt}\selectfont 
\centering
{\setlength\tabcolsep{4pt}
\begin{ThreePartTable}
\begin{TableNotes}
  \item[*] Detected by the overlapping region of the adjacent mosaic.
  \item[**] Detected by the overlapping region of two or more adjacent mosaics.
  \item[!] Detected by the overlapping region between Vela$-$\hi{} and Vela$-$SMGPS. 
\end{TableNotes}
\begin{longtable}{ccrrccccrrrrrl}
    \caption{Vela$-$\hi{} full catalogue: \hi{} data and derived properties.}
	\label{tab:cat}\\
    \hline
    \hline
           \multicolumn{1}{c}{Name} & \multicolumn{1}{c}{Mosaic} &        \multicolumn{1}{c}{$l$} &      \multicolumn{1}{c}{$b$} & 
           \multicolumn{1}{c}{$S_{\rm peak}$} &
           \multicolumn{1}{c}{$S_{\rm int}$} &  \multicolumn{1}{c}{$\varepsilon_s$} &  \multicolumn{1}{c}{RMS} &  $V_\mathrm{hel}$ &  $w_{20}$ &  $w_{50}$ &  $M_{\mathrm{HI}}$ & \multicolumn{1}{c}{Flag} &                               \multicolumn{1}{c}{Note} \\
           
            J2000 & AXB  &        \multicolumn{1}{c}{deg} &      \multicolumn{1}{c}{deg} & \multicolumn{1}{c}{Jy} &  Jy \kms{} &  \multicolumn{1}{c}{Jy \kms{}} &  \multicolumn{1}{c}{\mJy{}} &  \kms &  \kms &  \kms &  $\log (M_{\rm HI}$/\msun) & 1/2  &   \\
           
           \multicolumn{1}{c}{(1)} & \multicolumn{1}{c}{(2)} &        \multicolumn{1}{c}{(3)} &      \multicolumn{1}{c}{(4)} &  (5) &  \multicolumn{1}{c}{(6)} &  \multicolumn{1}{c}{(7)} &  (8) &  (9) &  (10) &  (11) & \multicolumn{1}{c}{(12)} &                               \multicolumn{1}{c}{(13)} & \multicolumn{1}{c}{(14)} \\
    \midrule
    \endfirsthead
    \caption*{Table \ref{tab:full_cat} continued}\\
    \hline
    \hline
    \multicolumn{1}{c}{Name} & \multicolumn{1}{c}{Mosaic} &        \multicolumn{1}{c}{$l$} &      \multicolumn{1}{c}{$b$} & 
           \multicolumn{1}{c}{$S_{\rm peak}$} &
           \multicolumn{1}{c}{$S_{\rm int}$} &  \multicolumn{1}{c}{$\varepsilon_s$} &  \multicolumn{1}{c}{RMS} &  $V_\mathrm{hel}$ &  $w_{20}$ &  $w_{50}$ &  $M_{\mathrm{HI}}$ & \multicolumn{1}{c}{Flag} &                               \multicolumn{1}{c}{Note} \\
           
            J2000 & TXX  &        \multicolumn{1}{c}{deg} &      \multicolumn{1}{c}{deg} & \multicolumn{1}{c}{Jy} &  Jy \kms{} &  \multicolumn{1}{c}{Jy \kms{}} &  \multicolumn{1}{c}{\mJy{}} &  \kms &  \kms &  \kms &  $\log (M_{\rm HI}$/\msun) & 1/2  &   \\
           
           \multicolumn{1}{c}{(1)} & \multicolumn{1}{c}{(2)} &        \multicolumn{1}{c}{(3)} &      \multicolumn{1}{c}{(4)} &  (5) &  \multicolumn{1}{c}{(6)} &  \multicolumn{1}{c}{(7)} &  (8) &  (9) &  (10) &  (11) & \multicolumn{1}{c}{(12)} &                               \multicolumn{1}{c}{(13)} & \multicolumn{1}{c}{(14)} \\
    \midrule
    \endhead 
    \hline
    \insertTableNotes  
    \endfoot 
    \hline
    \endlastfoot

J081815-465710 &    A2D &  263.291 & -6.232 &  0.013 &   2.85 &      0.29 &       0.77 &   7804 &  251 &  211 &     9.93 &    1 &                                                    w. comp (J081828-471025) \\
 J081828-471025 &    A2D &  263.497 & -6.324 &  0.011 &   0.67 &      0.14 &       0.71 &   7822 &  153 &  105 &     9.30 &    1 &                                  B=$\ang{;;40}$, i, w.comp (J081815-465710) \\
 J081845-473700 &    A2D &  263.893 & -6.531 &  0.017 &   4.49 &      0.26 &       0.71 &   7511 &  337 &  300 &    10.09 &    1 &                                                                             \\
 J081859-483716 &    A2D &  264.757 & -7.055 &  0.013 &   0.64 &      0.12 &       0.98 &  18638 &  124 &   81 &    10.04 &    2 &                                                                           e \\
 J082034-464906 &    A2D &  263.401 & -5.829 &  0.002 &   0.89 &      0.13 &       0.70 &   2148 &   67 &   50 &     8.30 &    1 &                                                                             \\
 J082116-474921 &    A2D &  264.305 & -6.293 &  0.010 &   2.42 &      0.33 &       0.66 &  11824 &  405 &  363 &    10.22 &    1 &                                                                           i \\
 J082148-463756 &    A2D &  263.366 & -5.549 &  0.011 &   1.27 &      0.21 &       0.72 &   9704 &  197 &  155 &     9.77 &    1 &                                                                             \\
 J082210-473533 &    A2D &  264.198 & -6.040 &  0.011 &   1.86 &      0.19 &       0.67 &   7657 &  222 &  199 &     9.73 &    1 &                                                                   i, w.comp \\
\bottomrule
\end{longtable}
\end{ThreePartTable}
}
\end{landscape}
}
\clearpage
\twocolumn

\bsp	
\label{lastpage}
\end{document}